\documentclass[12pt]{article}      

\usepackage{amsmath}
\usepackage{epsfig}

\usepackage[T1]{fontenc}

\usepackage{graphicx}
\usepackage{pstricks,pst-node,pst-text,pst-3d}
\usepackage{graphics}

\title{ The Bell Based Super Coherent States. Uncertainty Relations, Golden Ratio and Fermion-Boson Entanglement}

\author{ Oktay K Pashaev and Aygul Kocak\\Department of Mathematics\\ Izmir Institute of Technology \\ Urla-Izmir, 35430, T\"urkiye}

\begin{document}
\newcommand{\be}{\begin{equation}}
\newcommand{\ee}{\end{equation}}
\newcommand{\bea}{\begin{eqnarray}}
\newcommand{\eea}{\end{eqnarray}}
\newcommand{\disp}{\displaystyle}
\newcommand{\la}{\langle}
\newcommand{\ra}{\rangle}

\newtheorem{thm}{Theorem}[subsection]
\newtheorem{cor}[thm]{Corollary}
\newtheorem{lem}[thm]{Lemma}
\newtheorem{prop}[thm]{Proposition}
\newtheorem{definition}[thm]{Definition}
\newtheorem{rem}[thm]{Remark}
\newtheorem{prf}[thm]{Proof}

\maketitle


\begin{abstract}
 The set of maximally fermion-boson entangled Bell super-coherent states is introduced.  A superposition of these states with separable bosonic coherent states, represented by  points on the super-Bloch sphere,
we call the Bell based super-coherent states. 
Entanglement of bosonic and fermionic degrees of freedom in these states  is studied by using displacement bosonic operator.
It  acts on the super-qubit reference state, representing superposition of the zero and the one super-number states, forming computational  basis super-states.
We show that the states are completely characterized by displaced Fock states, as a superposition with non-classical, the photon added coherent states, and the entanglement is independent of coherent state parameter $\alpha$ and of the time evolution.
In contrast to never orthogonal Glauber coherent states, our entangled super-coherent states can be orthogonal. 
 The uncertainty relation in the states  is monotonically growing function of the concurrence and
for entangled states we get non-classical quadrature squeezing and
representation of uncertainty
by ratio of two Fibonacci numbers. The  sequence of concurrences,
 and corresponding
uncertainties $\hbar F_n/F_{n+1}$, in the limit $n \rightarrow \infty $, convergent to the  Golden ratio uncertainty  $\hbar/\varphi$, where $\varphi = \frac{1 + \sqrt{5}}{2}$ is found.
\end{abstract}


\section{Introduction}

It is well known that in contrast to coherent states as maximally classical states \cite{klauder},  the so called photon added states (PAC)\cite{Agarwal}, \cite{Francis} (described by adding single photon to coherent state)  are non-classical states with several specific properties as quadrature squeezing, sub-Poissonian distribution etc. \cite{Francis}.
As non-classical states, they attract interest in applications to quantum
sensing, quantum information processing, quantum
state engineering and probing fundamental properties of quantum mechanics. As non-Gaussian states, which could represent realistic experiments,
they become an object of research in continuous variable quantum information theory \cite{Francis}.
To
implement transition from classical to non-classical states, the superposition of coherent states with PAC states where studied theoretically and experimentally (see \cite{Zavatta} and references therein). Denoting the Glauber coherent state as $|0,\alpha\rangle = D(\alpha) |0\rangle$,
the displaced Fock state $|1,\alpha\rangle = D(\alpha) |1\rangle$ can be represented  as  such type superposition $|1,\alpha \rangle = a^\dagger |0,\alpha\rangle + \bar\alpha |0,\alpha \rangle$.
Then, the generic superposition of the coherent state and the PAC state appears  as displaced state $a |0,\alpha \rangle + b |1,\alpha\rangle = D(\alpha) (a |0\rangle + b |1\rangle)$
of the one qubit state  $a |0\rangle + b |1\rangle$. For $b \neq 0$, the state
is linearly independent on $|0,\alpha\rangle$, so that if we combine the states  as two component spinor of Fock states, it becomes descriptive of entangled fermion-boson states of supersymmetric quantum oscillator. This allows us  to study transition from classical to nonclassical states in framework of supersymmetric quantum mechanics \cite{Cooper}, and its dependence on  entanglement between fermions and bosons.

The purpose of this
paper is to study such type of states as entangled supersymmetric coherent states.
By using displacement bosonic operator, the states are generated
by acting on a reference state, that in addition to traditional vacuum state $|\Psi_0\rangle = |0\rangle_b \otimes |0\rangle_f$, includes superposition with the one super-particle state. This superposition is naturally to call as the superqubit state.
If the usual qubit state is a superposition of
$|0\rangle$ and $|1\rangle$ computational states, as eigenstates of the number operator $N_f$, the super-qubit state we define as superposition
\begin{equation}
|\theta, \phi \rangle_S = \cos {\frac{\theta}{2}} |0\rangle_S + \sin \frac{\theta}{2} e^{i \phi} |1\rangle_S
\end{equation}
of super-computational states $|0\rangle_S$ and $|1\rangle_S$, as eigenstates of the super-number operator ${\cal N}$, which counts number of superparticles. The states are parametrized by coordinates on the unit sphere, which we call as the super-Bloch sphere. Contribution of superparticles to energy is the same and does not distinguish fermions from bosons, so that the superqubits are the  degenerate states, but with different level of fermion-boson entanglement. In this paper
we  work with superqubit state as a superposition of separable $|0\rangle_S$ state and maximally entangled  $|1\rangle_S$ state. For the last one we use the first pair of Bell states in fermion-boson basis. After applying displacement operator to the superqubit state we get the first pair of super-coherent states. The second pair of states is generated from the second pair of fermion-boson Bell states, being not exact eigenstates of the supernumber operator (but only in averages).

We show that the fermion-boson entanglement in super-coherent state is equal to the one in  the corresponding super-qubit reference state and does not depend on
displacement parameter $\alpha$. The  entangled super-coherent states describe
transition from classical to non-classical behavior in non-minimal form of uncertainty relations, quadrature squeezing and sub-Poissonian distribution. The entanglement of states is independent of time evolution
and in contrast to Glauber coherent states, which are never orthogonal, the super-coherent states can be orthogonal. Depending on value of concurrence, we have the full circle of the equidistant  maximally entangled states (C=1), orthogonal to the given one, and the pair of orthogonal antipodal states for arbitrary $0<C<1$. For $C =1$, three mutually orthogonal states, associated with equilateral triangle in complex plane are found. For separable states with $C =0$, no orthogonal states are possible. This shows that entanglement of bosons with fermions is required to have orthogonality of coherent states. The entanglement affects also uncertainty relations.
The coordinate-momentum uncertainty  for supercoherent states,  represented in terms of monotonically growing function of the concurrence,  allows us to relate the uncertainty with level of boson-fermion entanglement. For the states along equator of the super- Bloch sphere, with $C = \frac{1}{2}$ we find the representation of uncertainty
by ratio of two Fibonacci numbers $\hbar F_5/F_6$. Then, by using the sequence of concurrences $C_n = \sqrt{F_{n-2}/F_{n+1}}$, convergent to Golden ratio form of entanglement $C = \varphi^{-3/2}$ we obtain the sequence of uncertainties $\hbar F_n/F_{n+1}$, in the limit $n \rightarrow \infty $ convergent to the  Golden ratio uncertainty $\hbar/\varphi$.

Similarly to superposition of coherent states with  PAC states,  making the state non-classical, our supercoherent states show the quadrature squeezing - when uncertainty in $X$   variable is lower than $1/2$, by expense of  increasing uncertainty in $P$ variable,  bigger than $1/2$, and vice versa.  This result can improve the measurement limits in SUSY quantum oscillator and
applied in several fields as quantum communications and quantum sensing, quantum optics and information processing.

Here we briefly describe existing literature in the field (list of which never could be complete)  and the differences with our paper.
The supersymmetric coherent states were studied in several papers from different points of view.
  In \cite{aragone},   for  simplest  ( N = 1) supersymmetric generalization of the standard quantum mechanical harmonic oscillator,
the supersymmetric annihilation operator $A_1 = I_f \otimes a + f \otimes I_b$ was defined. The super-coherent states as the eigenstates of this operator
were determined by using Fock space expansion.
First differences with \cite{aragone} we have  in terminology and representation of supersymmetric states.
 In contrast to the paper, we use fermion number operator $N_f = diag (0,1)$ and it leads to that  we have opposite number
of fermions in the given states. Our choice is motivated by quantum computation and quantum information theory, where computational basis state $|0\rangle$ we associate with zero fermions,
while state $|1\rangle$ - with one fermion.
The same definition is used in \cite{Cooper}.
The definition of the fermionic and bosonic states in  paper \cite{aragone} also  is not consistent with their calculations, since averages and uncertainty relations in Eq.(24) from \cite{aragone},
definitely take form of bosonic relations, but are called by authors as the fermionic ones.
Another difference  is that in the paper \cite{aragone} the authors construct the coherent states by formal expansion in Fock states, without analysis of the states structure,
while in our case we use the displacement operator, acting on different
reference states (not just the vacuum one) as superqubit states, based on the fermion-boson Bell states.
 This approach, together with calculation of concurrence and von Neumann entropy,  simplifies much the calculations and clarifies meaning of uncertainty relations and the entanglement property. In addition, we find orthogonality of entangled coherent states, quadrature squeezing,  Fibonacci sequences and Golden uncertainty.
One more specific is that we have four different super-annihilation operators $A$, which include not only $f$ operator, annihilating $|0\rangle$ state, but also $f^\dagger$, annihilating $|1\rangle$ state.

The paper  \cite{Hussin} is working with  $A_0 = I_f \otimes a$, as another super-annihilation operator, trying to formulate three equivalent definitions of supercoherent states, similarly to the Glauber states.
 As we show in present paper,  this operator can not generate entangled states from the separable ones.
More general form of nonlinear super-annihilation operator were studied in \cite{kornbluth} by Fock space expansion.
The group-theoretical approach with Grassman variables to supercoherent states was subject of paper \cite{Fatyga}, and Nieto in \cite{Nieto} proposed interpretation of the Grassman coherent state as the photino, the superpartner of photon.

The influence of squeezing operator on uncertainty relation for SUSY oscillator was subject of paper  \cite{orszag}.
It was shown that  unitary displacement operator for supercoherent states could be in the form $I \otimes D(\alpha)$ only,  and this  is exactly the one we are using in present paper, though the authors strangely mention that this pure $bosonic$ operator creates coherent states in $fermionic$ sector, which is misleading  and is related with using wrong definition in [1].
Specific form of displacement operator as the translation operator was used in \cite{Zypman} for supersymmatric displaced number states. All above papers are not discussing the influence of fermionic-bosonic entanglement on super-coherent states. In fact,
not many paper are devoted to  entanglement of bosons with fermions in SUSY.
Finite supersymmetry transformations and highly entangled combinations of bosons and fermions, invariant under supertranslations were worked out in \cite{Iliyeva}. In \cite{Laba},  by exploring
the Pauli Hamiltonian, entanglement of spin variables with continuous variables of electron in
uniform magnetic field, which exhibit SUSY was examined. They
determine the concurrence by the mean value of spin and  calculated it explicitly for SUSY quantum mechanical
states.
 The entanglement entropy in Gaussian states, related by SUSY is subject of discussion in paper of Jonnson \cite{Jonsson}.
In paper \cite{Motamed} entanglement of generalized supercoherent states with nonlinearly extended operator A were studied, but the authors
just copied determinant formula for concurrence of two qubit states without justification
for boson-fermion entanglement, which can not be considered as the correct one. This is why they found that concurrence depends on complex parameter of the bosonic coherent state $z$
($\alpha$ in our case), in contrast with our results of independence. For $\theta =0$ the concurrence in the paper depends on $z$ and angle $\phi$, which is different from our result.

The paper is organized as follows.
In Section 2, for  bipartite fermionic and bosonic system, the concurrence and the von Neumann entropy for reduced density matrix are
calculated  in  $n \rightarrow \infty$
limit of a qubit- n-qudit state. 
 In Section 3, we introduce supersymmetric qubit states, as superposition of zero and one superparticle states, by choosing the last one in fermionic-bosonic  Bell form.
 For geometric  representation of these states we introduce  the super-Bloch sphere.
In Section 4, we introduce four types of super-coherent states, as displaced super-qubit reference states,  based on fermion-boson Bell states.
In Section 5, we show that entanglement of supercoherent states, Section 5.1,  is the same as in the corresponding superqubit states, Section 5.2. In Section 5.3, condition for orthogonality of super-coherent states is found and the set of orthogoal supercoherent states is decribed. In Section 6 we show time independence of entanglement. The uncertainty relations and entanglement on super-Bloch sphere are subject of Section 7. In Section 7.1 we show squeezing  of quadrature as non classical property of our states.
In Section 7.2 we find the Golden ratio  uncertainty relation and link it  with Fibonacci numbers and Fibonacci sequence of concurrences. In Conclusions we discuss our results and future work.

\section{Fermion-Boson States}

Let ${f}$ and ${f}^{\dag}$ are fermionic annihilation and creation operators,
$
{f} {f}^{\dag}+{f}^{\dag}{f}={\textrm{I}}.
$
The eigenstates $|0\rangle_f$ and $|1\rangle_f$  of ${N}_{f}={f}^{\dag}{f}$, corresponding to fermionic numbers $n_{0}=0$ and $n_{1}=1$
 we denote as the qubit basis states.
 Normalized linear combination of these states determines the qubit unit of quantum information
\begin{equation}
|\theta, \phi \rangle = \cos {\frac{\theta}{2}} |0\rangle + \sin \frac{\theta}{2} e^{i \phi} |1\rangle,
\end{equation}
parametrized by points on the Bloch sphere $S^2$: $0 \le \theta \le \pi$, $0 \le \phi \le 2\pi$.

\subsection{Fermion-Boson Qubit States}

To work with fermionic and bosonic states, we first introduce the qubit-qudit state from Hilbert space $H_f \otimes H_n$, and then, to have the Fock space for  bosons  we take the limit
$n \rightarrow \infty$. The qudit state is determined by basis vectors $|0\rangle, |1\rangle, ..., |n-1 \rangle$ as computational states, and
generic qubit-qudit state
is 
\begin{eqnarray}
|\Psi\rangle = \sum^{n-1}_{k=0} c_{0 k} |0\rangle_f \otimes |k\rangle +
\sum^{n-1}_{k=0} c_{1 k} |1\rangle_f \otimes |k\rangle.
\end{eqnarray}
The state can be rewritten in two different forms. The first one
\begin{eqnarray}
|\Psi\rangle = |0\rangle_f \otimes |\psi_0\rangle + |1\rangle_f \otimes |\psi_1\rangle = \left(
          \begin{array}{c}
            |\psi_0\rangle  \\
            |\psi_1 \rangle        \\
          \end{array}
        \right),
\end{eqnarray}
represents it in terms of the pair of one qudit states
\begin{eqnarray}
|\psi_0 \rangle = \sum^{n-1}_{k=0} c_{0 k} |k \rangle, \,\,\,\,\,|\psi_1 \rangle = \sum^{n-1}_{k=0} c_{1 k} |k \rangle.
\end{eqnarray}
In the second one,
\begin{eqnarray}
|\Psi \rangle = |\varphi_0\rangle_{f} \otimes |0\rangle + |\varphi_1\rangle_{f} \otimes |1\rangle + ... + |\varphi_{n-1}\rangle_{f} \otimes |n-1\rangle = \sum^{n-1}_{l=0} |\varphi_l\rangle_{f} \otimes |l\rangle
\end{eqnarray}
it is given by  $n$, the one qubit states $|\varphi_{l}\rangle$, $l=0,...,n-1$, defined as
\begin{eqnarray}
|\varphi_{l}\rangle = \left(
          \begin{array}{c}
            c_{0 l}  \\
            c_{1 l}        \\
          \end{array}
        \right) = c_{0 l} |0\rangle_f + c_{1 l} |1\rangle_f .
\end{eqnarray}

Now, we send dimension of the qudit state, $n \rightarrow \infty$, so that the space of states $H_n$ becomes  the Fock space $H_b$,
and the computational basis of qudit states transforms to Fock number states  $|k \rangle_\infty \equiv |k \rangle $, $k=0,1,2,...$. The fermionic-bosonic  basis states are formed
by tensor product of fermionic (qubit) states with Fock states, $|0\rangle \otimes |k\rangle$, and $|1\rangle \otimes |k \rangle$, $k=0, 1, 2,...$ and 
for arbitrary state
\begin{equation}
|\Psi \rangle = \sum^\infty_{k=0} c_{0 k} |0\rangle \otimes |k\rangle +  \sum^\infty_{k=0} c_{1 k} |1\rangle \otimes |k\rangle, \label{BFstate}
\end{equation}
from $H_f \otimes H_b$ Hilbert space we have  two representations. The first one is
\begin{eqnarray}
|\Psi\rangle = |0\rangle \otimes |\psi_0\rangle + |1\rangle \otimes |\psi_1\rangle = \left(
          \begin{array}{c}
            |\psi_0\rangle  \\
            |\psi_1 \rangle        \\
          \end{array}
        \right), \label{firstrepresentation}
\end{eqnarray}
where two bosonic states
\begin{eqnarray}
|\psi_0 \rangle = \sum^{\infty}_{k=0} c_{0 k} |k \rangle, \,\,\,\,\,|\psi_1 \rangle = \sum^{\infty}_{k=0} c_{1 k} |k \rangle. \label{bosonicstates}
\end{eqnarray}
are vectors  in the Fock space.
The second representation
\begin{equation}
|\Psi \rangle = |\varphi_0\rangle \otimes |0\rangle + |\varphi_1\rangle \otimes |1\rangle + ... + |\varphi_{n}\rangle \otimes |0\rangle + ... = \sum^{\infty}_{n=0} |\varphi_n\rangle \otimes |n\rangle ,
\end{equation}
is determined by infinite set of qubits $|\varphi_n \rangle$, $n=0,1,2,...$,
defined as
\begin{eqnarray}
|\varphi_{n}\rangle = \left(
          \begin{array}{c}
            c_{0 n}  \\
            c_{1 n}        \\
          \end{array}
        \right) = c_{0 n} |0\rangle + c_{1 n} |1\rangle .\label{infinitequbits}
\end{eqnarray}

\subsection{Entanglement of Fermion-Boson States}

The fermionic-bosonic  state from $H_f \otimes H_b$ is separable if
$
|\Psi \rangle = |\Phi\rangle_f \otimes |\Xi\rangle_b,
$
where $|\Phi\rangle_f$ is the one qubit or the fermionic state, and $|\Xi\rangle_b$ is bosonic state from the Fock space.
If the state $|\Psi \rangle$ is not separable, then it is  entangled.

\begin{prop}
The state (\ref{BFstate}) is separable if and only if
in representation (\ref{firstrepresentation}) two Fock states (\ref{bosonicstates}) are linearly dependent, $|\psi_0 \rangle = \lambda |\psi_1 \rangle$.
\end{prop}

If these states are linearly  independent, the state (\ref{BFstate}) is entangled.
To find the level of entanglement for the generic pure state (\ref{BFstate}), we calculate the reduced density matrices.
For normalized state in (\ref{firstrepresentation})
the density matrix is
\begin{equation}
\rho = | \Psi \rangle \langle \Phi | = \left(
 \begin{array}{cc}
  | \psi_0 \rangle \langle \psi_0| &     | \psi_0 \rangle \langle \psi_1|          \\ \\
    | \psi_1 \rangle \langle \psi_0| &  | \psi_1 \rangle \langle \psi_1| \\
	\end{array} \right),
\end{equation}
and due to normalization condition,
$
\sum_{n=0}^\infty (|c_{0 n}|^2 + |c_{1 n}|^2) =1
$,
\begin{equation}
tr \rho = \langle \psi_0 | \psi_0 \rangle + \langle \psi_1 | \psi_1 \rangle = 1. \label{tracerho}
\end{equation}
 For the reduced bosonic density matrix
\begin{equation}
\rho_b = tr_f \,\rho = | \psi_0 \rangle \langle \psi_0| + | \psi_1 \rangle \langle \psi_1|
\end{equation}
 we find
\begin{equation}
tr \rho^2_b = \langle \psi_0 | \psi_0 \rangle^2 + \langle \psi_1 | \psi_1 \rangle^2 + 2 |\langle \psi_0 | \psi_1 \rangle|^2,
\end{equation}
and for the fermionic one
\begin{equation}
\rho_f = tr_b \,\rho = \sum^\infty_{n=0} |\varphi_n \rangle \langle \varphi_n | \label{reduceddensitymatrixf}
\end{equation}
the expression is
\begin{equation}
tr \rho^2_f = \sum^\infty_{n=0} \sum^\infty_{m=0}    |\langle \varphi_n | \varphi_m \rangle|^2.
\end{equation}
As easy to check by direct computation,  it coincides with the bosonic one,  so that $tr \rho^2_b = tr \rho^2_f$.
The first one we rewrite in the form
\begin{equation}
tr \rho^2_b = (\langle \psi_0 | \psi_0 \rangle + \langle \psi_1 | \psi_1 \rangle)^2 - 2 (\langle \psi_0 | \psi_0 \rangle \langle \psi_1 | \psi_1 \rangle - \langle \psi_0 | \psi_1 \rangle \langle \psi_1 | \psi_0 \rangle)
\end{equation}
and by taking into account the squared equation (\ref{tracerho}), we get
\begin{equation}
1 - tr \rho^2_b = 2 \left|
 \begin{array}{cc}
  \langle\psi_0 | \psi_0 \rangle &        \langle\psi_0 | \psi_1 \rangle       \\ \\
     \langle\psi_1 | \psi_0 \rangle &  \langle\psi_1 | \psi_1 \rangle  \\
	\end{array} \right|.     \label{difference}
\end{equation}
Deviation of the trace from unity gives a simplest characteristics of the level of entanglement.
It is known as the linear entropy (see \cite{LE} and references there), appearing in the linear approximation of the von Neumann entropy.
In following, for this difference we introduce definition of the concurrence $C$  in the determinant form, normalized as
for the two qubit states
\cite{ParlakPashaev}.
\begin{definition} The concurrence $C$ of a pure fermion-boson state is defined by reduced density matrix  $\rho_f$ (or $\rho_b$) as
the number
\begin{equation}
C = \sqrt{2} \sqrt{1- tr \rho_f^2},
\end{equation}
satisfying
\begin{equation}
tr \rho_f^2 + \frac{1}{2} C^2 =1. \label{concurrence}
\end{equation}
\end{definition}
From (\ref{concurrence}) and (\ref{difference}) we find the concurrence square as determinant of the Hermitian inner product metric
$g_{ij} = \langle\psi_i | \psi_j \rangle$, (the Gram determinant), of two vectors $(i,j = 0,1)$ in Fock space,
\begin{equation}
C^2 = 4\left|
 \begin{array}{cc}
  \langle\psi_0 | \psi_0 \rangle &        \langle\psi_0 | \psi_1 \rangle       \\ \\
     \langle\psi_1 | \psi_0 \rangle &  \langle\psi_1 | \psi_1 \rangle  \\
	\end{array} \right|,
\end{equation}
and for the generic quantum state (\ref{BFstate}),
\begin{equation}
C = 2 \,\sqrt{det \left(
 \begin{array}{cc}
  \langle\psi_0 | \psi_0 \rangle &        \langle\psi_0 | \psi_1 \rangle       \\ \\
     \langle\psi_1 | \psi_0 \rangle &  \langle\psi_1 | \psi_1 \rangle  \\
	\end{array} \right) }. \label{genericconcurrence}
\end{equation}
Due to relation
\begin{equation}
tr \rho^2_f = 1 - \sum^\infty_{n=0} \sum^\infty_{m=0} \left|
 \begin{array}{cc}
  \langle\varphi_n | \varphi_n \rangle &        \langle\varphi_n | \varphi_m \rangle       \\ \\
     \langle\varphi_m | \varphi_n \rangle &  \langle\varphi_m | \varphi_m \rangle  \\
	\end{array} \right|,
\end{equation}
the concurrence can be represented also in another form
\begin{equation}
C^2 =2 \sum^\infty_{n=0} \sum^\infty_{m=0} \left|
 \begin{array}{cc}
  \langle\varphi_n | \varphi_n \rangle &        \langle\varphi_n | \varphi_m \rangle       \\ \\
     \langle\varphi_m | \varphi_n \rangle &  \langle\varphi_m | \varphi_m \rangle  \\
	\end{array} \right|.
\end{equation}

By using explicit form of the one qubit states (\ref{infinitequbits}) it can be rewritten as an infinite sum of modulus squares of all $2\times 2$ minors of the coefficient matrix $c_{n m}$,
\begin{equation}
C^2 =4 \sum^\infty_{0=n < m} \left| \left|
 \begin{array}{cc}
  c_{0 n} &   c_{0 m}      \\ \\
     c_{1 n} &  c_{1 m} \\
	\end{array} \right|\right|^2.
\end{equation}
This way we get two equivalent representations for the concurrence.
\begin{prop}
For generic normalized fermion-boson state (\ref{BFstate}) from Hilbert space $H_f \otimes H_b$, the concurrence is equal
\begin{equation}
C= 2 \,\sqrt{det \left(
 \begin{array}{cc}
  \langle\psi_0 | \psi_0 \rangle &        \langle\psi_0 | \psi_1 \rangle       \\ \\
     \langle\psi_1 | \psi_0 \rangle &  \langle\psi_1 | \psi_1 \rangle  \\
	\end{array} \right)}
=2 \,\sqrt{ \sum^\infty_{0=n < m} \left| \left|
 \begin{array}{cc}
  c_{0 n} &   c_{0 m}      \\ \\
     c_{1 n} &  c_{1 m} \\
	\end{array} \right|\right|^2 }. \label{Cformula}
\end{equation}
\end{prop}

\begin{cor}
The determinant of $2\times 2$ inner product metric in Fock space can be represented by an infinite sum of modulus squares of minors of the infinite matrix from
coefficients $c_{n m}$ of the state (\ref{BFstate}),
\begin{eqnarray}
det \left(
 \begin{array}{cc}
  \langle\psi_0 | \psi_0 \rangle &        \langle\psi_0 | \psi_1 \rangle       \\ \\
     \langle\psi_1 | \psi_0 \rangle &  \langle\psi_1 | \psi_1 \rangle  \\
	\end{array} \right)  =  \sum^\infty_{0=n < m} \left| \left|
 \begin{array}{cc}
  c_{0 n} &   c_{0 m}      \\ \\
     c_{1 n} &  c_{1 m} \\
	\end{array} \right|\right|^2.
\end{eqnarray}
\end{cor}
By using the definition and above expressions for the concurrence, now we calculate entanglement in fermion-boson system by the von Neumann entropy.

\begin{prop}
The entanglement, as the value of the von Neumann entropy
\begin{equation}
E_f = - tr ( \rho_f \, \log_2 \rho_f  ) \label{VNentropy}
\end{equation}
for $\rho_f$ in (\ref{reduceddensitymatrixf}) is
\begin{equation}
E_f = - \frac{1 + \sqrt{1- C^2}}{2} \log_2 \frac{1 + \sqrt{1- C^2}}{2} - \frac{1 - \sqrt{1- C^2}}{2} \log_2 \frac{1 - \sqrt{1- C^2}}{2} \label{entanglement}
\end{equation}
where the concurrence $C$ is given by  (\ref{Cformula}). The value of concurrence is bounded between $0 \le C \le 1$.
\end{prop}
\begin{prf}
The characteristic equation for matrix $\rho_f$,
\begin{equation}
\lambda^2 - \lambda + det\, \rho_f =0
\end{equation}
has two real eigenvalues
\begin{equation}
\lambda_{1,2} = \frac{1}{2} \pm \sqrt{\frac{1}{4} - \det\, \rho_f},
\end{equation}
where the determinant of $\rho_f$ can be expressed by the concurrence as
\begin{equation}
det\, \rho_f =  \sum^\infty_{0=n < m} \left| \left|
 \begin{array}{cc}
  c_{0 n} &   c_{0 m}      \\ \\
     c_{1 n} &  c_{1 m} \\
	\end{array} \right|\right|^2 = \frac{1}{4} C^2
\end{equation}

\end{prf}
We note that for the fermion-boson system the entanglement $E_f$ is function of $C$ only, though the last one  includes infinite sum of modulus squares of $2\times 2$ minors.

 \section{Supersymmetric Qubit States}

\subsection{The Supersymmetric Harmonic Oscillator}

The supersymmetric(SUSY) harmonic oscillator is a composition of fermionic and bosonic harmonic oscillators with equal frequencies \cite{Cooper},
\begin{eqnarray}
{\textrm{H}}={\textrm{H}}_{b}+ {\textrm{H}}_{f}
=\frac{\omega}{2}\left\{{a},{a}^{\dag}\right\}+\frac{\omega}{2}\left[{f}^{\dag},{f}\right]
=\omega {\cal N}.
\end{eqnarray}
 Here, the super-number operator ${\cal N}$
\begin{equation}
{\cal N} = {I}_{f}\otimes {N} + {N}_{f}\otimes {I}_{b}=\left(
                                                                                           \begin{array}{cc}
                                                                                             {N} & 0 \\
                                                                                             0 & {N}+I_b \\
                                                                                           \end{array}
                                                                                         \right) \, ,\label{supernumber}
\end{equation}
has eigenstates $|n_{b},n_{f}\rangle=|n_{f}\rangle\otimes|n_{b}\rangle,$  where
 $n_{b}=0,1,2,...$ and $n_{f}=0,1$ are eigenvalues of bosonic and fermionuc number operators correspondingly, $N |n_{b}\rangle = n_b |n_{b}\rangle$, $N_f |n_{f}\rangle = n_f |n_{f}\rangle$.  It counts the total number of fermions and bosons $n = n_{b}+n_{f}$ in state $|n_{b},n_{f}\rangle$.
 The eigenstates $|0\rangle \otimes |n\rangle$, and  $|1\rangle \otimes |n-1\rangle$
 have the energy $E_{n}=n\omega , n>0$ and $E_{0}=0$, for $n=0.$ This shows that fermionic and bosonic quanta have the same energy $\omega$, and the states have the same number $n$ of supersymmetric boson-fermion quanta (super-particles or super-quanta). The difference between states is the number of fermions, which is zero in the first case (pure bosonic state) and is one in the second case. Moreover, an arbitrary
superposition of these two states is also state with $n$ super-quanta, which after normalization can be written as the super-number state
\begin{equation}
|n, \theta, \phi \rangle=\cos \frac{\theta}{2}\left(
  \begin{array}{c}
    |n\rangle\\
    0 \\
  \end{array}
\right)+\sin \frac{\theta}{2} e^{i \phi}\left(
               \begin{array}{c}
                 0 \\
                 |n-1\rangle \\
               \end{array}
             \right)
               \quad .  \label{supernumberstate}
\end{equation}

This shows that the
energy levels with $n$ super-quanta $E_{n>0}=n\omega$ are double degenerate with arbitrary $0 \le \theta \le \pi$, $0 \le \phi \le 2 \pi$. For $n=0$ the state
$|\Psi_{0} \rangle = |0\rangle_f \otimes |0\rangle_b$
 is the ground state with $E_{0}=0.$
The super-number state (\ref{supernumberstate}) contains $n$ super-quanta, ${\cal N} |n, \theta, \phi \rangle = n |n, \theta, \phi \rangle$, in superposition of the zero fermionic state
$|0\rangle_f \otimes |n\rangle$ and the one fermionic state $|1\rangle_f \otimes |n-1\rangle$. For  this superposition,
 the probabilities do not depend on $n$ and are equal
\begin{equation}
\langle n, \theta, \phi |P_0 |n, \theta, \phi \rangle = \cos^2 \frac{\theta}{2} \equiv p_0, \,\,\,\,
\langle n, \theta, \phi |P_1 |n, \theta, \phi \rangle = \sin^2 \frac{\theta}{2} \equiv p_1,
\end{equation}
where projection operators are
$P_0 = (|0 \rangle \langle 0 |) \otimes I_b $, and $P_1 = (|1 \rangle \langle 1 |) \otimes I_b $.
This allows us to represent  the  super-qubit state (\ref{supernumberstate}) as a state on the Bloch type sphere, which is natural to call as the super-Bloch sphere, where the north pole of the sphere $\theta =0$, corresponds to the zero fermion state
and the south pole $\theta = \pi$ to the one fermion state. The states along the equator $\theta = \frac{\pi}{2}$ are in maximally random superposition  of these states. This way we have geometrical description of degeneracy of the $n$ super-quanta state in terms of the super-Bloch sphere.

\subsection{Entanglement of Super-Number States}
To evaluate level of entanglement between bosons and fermions  in  the super-number states (\ref{supernumberstate}), we  use the reduced density matrix method.
The density matrix for the pure state (\ref{supernumberstate}) is equal
\begin{equation}
\rho_n = |n, \theta, \phi \rangle \langle n, \theta, \phi |= \left(
\begin{array}{cc}
\cos^2 \frac{\theta}{2}\, |n \rangle \langle n | & \cos \frac{\theta}{2} \sin \frac{\theta}{2} e^{-i\phi}  |n \rangle \langle n-1 | \\
 & \\
   \cos \frac{\theta}{2} \sin \frac{\theta}{2} e^{i\phi}  |n-1 \rangle \langle n|   & \sin^2 \frac{\theta}{2} \,|n-1 \rangle \langle n-1 |\\
\end{array}
 \right)
\end{equation}
It satisfies $tr \rho_n =1$, $tr \rho_n^2 =1$.
By taking partial trace of $\rho_n$ according to  fermionic states we get the reduced bosonic density matrix
\begin{equation}
\rho_b = tr_f \,\rho_n = \sin^2 \frac{\theta}{2}\, |n-1 \rangle \langle n-1 | + \cos^2 \frac{\theta}{2} \,|n \rangle \langle n |,
\end{equation}
as an infinite dimensional matrix with only two nonzero diagonal terms,  $\sin^2 \frac{\theta}{2}$ and $\cos^2 \frac{\theta}{2}$ at positions $n$ and $n+1$, correspondingly.
 The partial trace according to  bosonic states gives  fermionic density matrix as $2\times 2$ diagonal matrix
\begin{equation}
\rho_f = tr_b \,\rho = \cos^2 \frac{\theta}{2} \,|0 \rangle \langle 0 | + \sin^2 \frac{\theta}{2} \,|1 \rangle \langle 1 | .
\end{equation}
For  trace of the square of both reduced density matrices we get
\begin{equation}
tr \rho_f^2 = tr \rho_b^2 = 1 - \frac{1}{2} \sin^2 \theta.
\end{equation}
Then, by using formula (\ref{concurrence}) we obtain that the reduced bosonic, as well as fermionic state is mixed and the generic state $|n, \theta, \phi \rangle$ is entangled with concurrence
\begin{equation}
C = \sin \theta. \label{concurrenceNsuper}
\end{equation}
It is bounded $0 \le C \le 1$ and does not dependent on $n$.  The north pole state $|n, \theta =0, \phi \rangle$ (n-bosons state), and the south pole state $|n, \theta =\pi, \phi \rangle$ (n-1 bosons and one fermion state) are separable for any $n$, and correspond to $C =0$.
Contrary, the states along the equator on super-Bloch sphere, $|n, \theta =\frac{\pi}{2}, \phi \rangle$ with the concurrence $C = 1$  are maximally entangled states.
The general form of these states is
\begin{equation}
|n, \frac{\pi}{2}, \phi \rangle = \frac{1}{\sqrt{2}} (|0\rangle \otimes |n\rangle + e^{i\phi} |1\rangle \otimes |n-1 \rangle ). \label{maxentangled}
\end{equation}

\subsection{Fermion-Boson Bell States}

For $n=1$ we have the maximally entangled states
\begin{equation}
|L_\phi \rangle \equiv
|1, \frac{\pi}{2}, \phi \rangle  = \frac{1}{\sqrt{2}} (|0\rangle_f |1\rangle_b + e^{i\phi}
|1\rangle_f |0\rangle_b),
\end{equation}
giving the fermion-boson analog of the Bell states ($\phi =0, \pi$),
\begin{equation}
|L_\pm \rangle \equiv |1, \frac{\pi}{2}, \pm \rangle=\frac{1}{\sqrt{2}} (|0\rangle_f |1\rangle_b \pm
|1\rangle_f |0\rangle_b). \label{Lstates}
\end{equation}

\begin{definition}
The states with $n$-superparticles
\begin{equation}
|n, \pm \rangle \equiv |n, \frac{\pi}{2}, \pm \rangle  =\frac{1}{\sqrt{2}} (|0\rangle_f |n\rangle_b \pm
|1\rangle_f |n-1\rangle_b), \label{generalizedBell}
\end{equation}
where $n=1,2,...$, we call as  generalized Bell states. For $n=1$ the states become just the fermionic-bosonic Bell states
$|L_\pm \rangle$ as in (\ref{Lstates}).
\end{definition}
The infinite set of these states is maximally entangled, $C =1$, for any positive integer $n$ and
satisfies orthonormality conditions
\begin{equation}
  \langle m, \pm    |n, \pm \rangle = \delta_{n,m}, \hskip1cm   \langle m, \mp   |n, \pm \rangle = 0.\hskip0.5cm                             n,m =1,2,...
	\label{generalBellorthogonal}
\end{equation}

In addition to the pair of Bell states (\ref{Lstates}) we introduce another pair of fermionic-bosonic Bell states
\begin{equation}
|B_\pm \rangle =\frac{1}{\sqrt{2}} (|0\rangle_f |0\rangle_b \pm
|1\rangle_f |1\rangle_b). \label{Bstates}
\end{equation}
The four Bell states (\ref{Lstates}),(\ref{Bstates}) are orthonormal
\begin{eqnarray}
\langle L_+| L_+\rangle &=& \langle L_-| L_-\rangle = 1, \hskip0.5cm \langle L_+| L_-\rangle = 0, \\
\langle B_+| B_+\rangle &=& \langle B_-| B_-\rangle =1, \hskip0.5cm \langle B_+| B_-\rangle = 0, \\
\langle B_\pm| L_\pm\rangle &=& \langle B_\pm| L_\mp\rangle =0,
\end{eqnarray}
and maximally entangled. It is noted that  in contrast with $|L_\pm\rangle$, the states (\ref{Bstates}) are not eigenstates of the supernumber operator. In fact,
states $|L_\pm \rangle$ are exact eigenstates of ${\cal N}$ with one superparticle $n=1$, ${\cal N} |L_\pm \rangle = |L_\pm \rangle$, while states $|B_\pm \rangle$ are not the eigenstates
and only the average number of superparticles in these states is one, $\langle B_\pm | {\cal N} |B_\pm \rangle =1$.

\subsection{The Bell based Super-qubit States}

In this paper, to create coherent states we will use the displacement operator ${\cal D}(\alpha)$ defined in (\ref{superdisplacement}). If this operator is acting on the vacuum $(n=0)$ state, $|\Psi_0\rangle = |0\rangle_f \otimes |0\rangle_b$,
annihilated by operator $A_0 = I_f \otimes a$, so that, $A_0 |\Psi_0 \rangle =0$, the corresponding coherent state as the eigenstate of this operator, would be separable. Another state, annihilated by this operator $|\Psi_1\rangle =  |1\rangle_f \otimes |0\rangle_b$, is the one particle state with $n=1$, and it is also separable. Moreover, any superposition  of these two, the vacuum and one particle states,
$\alpha (|0\rangle_f \otimes |0\rangle_b) + \beta (|1\rangle_f \otimes |0\rangle_b) = |0\rangle_f (\alpha |0\rangle_b +\beta |1\rangle_b)$ is separable.
To create the entangled fermionic-bosonic coherent state, instead of this, we have to
choose the reference state  as the entangled state. To proceed  in this direction, we first describe the maximally entangled states ($C=1$) and then take superposition of these states with the
separable ones ($C=0$). This way we get entangled states, depending on the concurrence parameter $C$ and implementing transition from separable to maximally entangled state.
The natural choice for maximally entangled states is the set of four fermionic-bosonic Bell states (\ref{Lstates}), (\ref{Bstates}).
Due to entanglement of bosons with fermions, these states are not annihilated by pure bosonic annihilation operator $A_0$ and require a mixture of bosonic and fermionic operators. In fact, for every Bell state we have its own annihilation operator, which in addition to bosonic annihilation operator $a$ includes
the fermionic annihilation or creation operators, $f$ and $f^\dagger$. We define four operators
\begin{eqnarray}
A_{\pm 1} &=& \left(
   \begin{array}{cc}
   {a} & \pm 1 \\
   0 & {a} \\
	\end{array} \right)   =  I_f \otimes a \pm f \otimes I_b,                                                            \label{A1} \\
	A^T_{\pm 1} &=& \left(
   \begin{array}{cc}
   {a} & 0 \\
   \pm 1 & {a} \\
	\end{array} \right)   =  I_f \otimes a \pm f^\dagger \otimes I_b ,                                                           \label{A2}
\end{eqnarray}
annihilating the following Bell states
\begin{eqnarray}
A_1 |L_{-}\rangle &=& 0, \hskip1cm A_{-1} |L_{+}\rangle =0, \\
A^T_1 |B_{-}\rangle &=& 0, \hskip1cm A^T_{-1} |B_{+}\rangle =0,
\end{eqnarray}
and acting as quantum gates, transforming the states to each other,
\begin{eqnarray}
A_1 |B_\pm \rangle &=& \pm |L_+\rangle,  \hskip1cm
A_{-1} |B_\pm \rangle = \mp |L_-\rangle, \\
A^T_1 |L_\pm \rangle &=&  |B_+\rangle, \hskip1cm
A^T_{-1} |L_\pm \rangle =  |B_-\rangle.
\end{eqnarray}
It is noted that the above supersymmetric annihilation operators include also creation operator $f^\dagger$. It should be  not surprising, since
action of this operator on one fermion state gives zero $f^\dagger |1\rangle_f = 0$. This is why, the set of the annihilation operators
become more reach and it is valid for any two level system or any qubit state.

The first pair of states $|L_\pm \rangle$ can be generated from the vacuum state $|\Psi_0\rangle$  and vice versa
\begin{equation}
|L_\pm \rangle = \frac{1}{\sqrt{2}} A^\dagger_{\pm1} |\Psi_0 \rangle,\hskip0.5cm    |\Psi_0\rangle = \frac{1}{\sqrt{2}}  A_{\pm1} |L_{\pm} \rangle,
\end{equation}
and the second pair of states $|B_\pm \rangle$ from the one fermion state $|\Psi_1\rangle$ by
\begin{equation}
|B_\pm \rangle = \pm\frac{1}{\sqrt{2}} (A^T_{\pm1})^\dagger |\Psi_1 \rangle,\hskip0.5cm    |\Psi_1\rangle = \pm \frac{1}{\sqrt{2}}    A^T_{\pm1} |B_{\pm} \rangle.
\end{equation}

The vacuum state is annihilated by two operators
\begin{equation}
A_{\pm 1} |\Psi_0 \rangle = 0,
\end{equation}
and it is orthogonal to the pair of Bell states $|L_+\rangle$ and $L_-\rangle$.
By taking superposition of the state with these Bell states we get two normalized reference states,
\begin{eqnarray}
|0, C, \phi \rangle_{L_\pm} &=& \sqrt{1-C} |\Psi_0 \rangle + \sqrt{C} e^{i\phi} |L_\pm \rangle,\label{refLpm}
\end{eqnarray}
which are annihilated by operators
\begin{equation}
A_{\mp 1}|0, C, \phi \rangle_{L_\pm} =0. \label{annihilateL}
\end{equation}
The states are parametrized by real number $C$, bounded between $0 \le C \le 1$.
It represents the concurrence, calculated from formula (\ref{Cformula}) and
showing the level of fermion-boson entanglement in the reference state.

The parametrization allows us to give two physical interpretations of concurrence $C$.
In the first one, it shows probability to measure the one superparticle state $|L_+\rangle$ or $|L_-\rangle$
\begin{equation}
C = \langle 0, C, \phi |P_1 |0, C, \phi \rangle = p_1
\end{equation}
in the superposition (\ref{refLpm})   of vacuum (zero superparticle state) and  $|L_\pm \rangle$ (one superparticle state).
The second meaning of $C$ is the average value of supernumber operator in the superposition state
\begin{equation}
C = _{L_\pm} \langle 0, C, \phi | {\cal N}|0, C, \phi \rangle_{L_\pm}.
\end{equation}

To calculate the second pair of reference states we notice that application of $f^\dagger$ operator on the vacuum state $|\Psi_0 \rangle$ generates one fermion state
\begin{equation}
f^\dagger |\Psi_0 \rangle = |\Psi_1 \rangle = \left(
                                   \begin{array}{c}
                                     0 \\
                                     |0 \rangle \\
                                   \end{array}
                                 \right),  \label{psi1}
\end{equation}
annihilated by operators
\begin{equation}
A^T_{\mp 1}|\Psi_1 \rangle =0
\end{equation}
and orthogonal to the second pair of Bell states $|B_+\rangle$ and $|B_- \rangle$.
Superposition of the state with  these Bell states gives another pair of reference states,
\begin{eqnarray}
|0, C, \phi \rangle_{B_\pm} &=& \sqrt{1-C} |\Psi_1 \rangle + \sqrt{C} e^{i\phi} |B_\pm \rangle,\label{refBpm}
\end{eqnarray}
which are annihilated by operators
\begin{equation}
A^T_{\mp 1}|0, C, \phi \rangle_{B_\pm} =0. \label{annihilateB}
\end{equation}

As a result,  we have constructed four, the Bell type reference states
\begin{eqnarray}
|0, C, \phi \rangle_{L_\pm} &=& \sqrt{1-C} |\Psi_0 \rangle + \sqrt{C} e^{i\phi} |L_\pm \rangle,\label{refLpm1} \\
|0, C, \phi \rangle_{B_\pm} &=& \sqrt{1-C} |\Psi_1 \rangle + \sqrt{C} e^{i\phi} |B_\pm \rangle,\label{refBpm1}
\end{eqnarray}
with the inner products
\begin{equation}
{_{L_{+}}}\langle 0, C, \phi | 0, C, \phi \rangle_{L_{-}} = 1 - C, \hskip0.5cm {_{B_{+}}}\langle 0, C, \phi | 0, C, \phi \rangle_{B_{-}} = 1 - C
\end{equation}
and corresponding  fidelity $F = (1-C)^2$, expressed in terms of the concurrence $C$.
The reference states are characterized by real number $C$, bounded as $0 \le C \le 1$,  and the angle $0 \le \phi \le 2\pi$ . This is why geometrically, every state
represents the point on surface of circular cylinder with radius one and height $C$. Another geometrical image of these states is associated with the unit disk $|z| \le 1$ in complex plane $z = \sqrt{C} e^{i\phi}$.
One more representation is given by points on unit sphere, parametrized  by two angles $0\le \theta \le \pi$ and $0\le\phi \le 2\pi$, related to concurrence by
$\sin \frac{\theta}{2} = \sqrt{C}$, $\cos \frac{\theta}{2} = \sqrt{1 - C}$.
\begin{definition}
The reference states
\begin{eqnarray}
|0, \theta, \phi \rangle_{L_\pm} &=& \cos \frac{\theta}{2} |\Psi_0 \rangle + \sin \frac{\theta}{2} e^{i\phi} |L_\pm \rangle,\label{refLpm2}
\end{eqnarray}
as superposition of zero super-particle state and one super-particle state are called the super-qubit states. Every state is represented by point on the unit sphere, which we call the super-Bloch sphere.
\end{definition}
The north pole of the sphere corresponds to separable vacuum state,
while the south pole to maximally entangled Bell state. Similarly to the usual qubit state, the north pole state $|\Psi_0\rangle \equiv |0\rangle_S$ is  $n=0$ superparticle state, ${\cal N} |0\rangle_S =0\,|0\rangle_S$, and the south pole
 state $|L_\pm\rangle \equiv |1\rangle_S$ is
$n=1$ superparticle state, ${\cal N} |1\rangle_S =1 \,|0\rangle_S$. However, the state is fermion-boson entangled and the computational  basis for this super-qubit state is made from $|0\rangle_S$ and $|1\rangle_S$ eigenstates of super-number operator ${\cal N}$.

The second pair of reference states is defined as
\begin{eqnarray}
|0, \theta, \phi \rangle_{B_\pm} &=& \cos \frac{\theta}{2}  |\Psi_1 \rangle + \sin \frac{\theta}{2} e^{i\phi} |B_\pm \rangle,\label{refBpm2}
\end{eqnarray}
but basis states are not eigenstates of ${\cal N}$ operator.
In next Section, by acting on these states by displacement operator  we construct four orthogonal super-coherent states.

\section{The Supersymmetric Coherent States}

To construct supersymmetric coherent state we follow the displacement operator approach.

 \subsection{Displacement Operator}

We introduce the bosonic displacement operator as the direct product
\begin{equation}
{\cal D}(\alpha) = \left(
                                                                                                            \begin{array}{cc}
                                                                                                              D(\alpha) & 0 \\
                                                                                                              0 & D(\alpha) \\
                                                                                                            \end{array} \right)
		= I_f \otimes D(\alpha) = I_f \otimes e^{\alpha a^\dagger - \bar\alpha a},			\label{superdisplacement}																																																	
\end{equation}
satisfying unitarity condition
${\cal D}(\alpha) {\cal D}^\dagger(\alpha) = I.$
Applying this operator to vacuum state $|\Psi_0 \rangle$ and the one fermion state (\ref{psi1}) we get corresponding supersymmetric coherent states
\begin{eqnarray}
{\cal D}(\alpha) |\Psi_0 \rangle = \left(
                                   \begin{array}{c}
                                     D(\alpha)|0\rangle \\
                                     0 \\
                                   \end{array}
                                 \right) = \left(
                                   \begin{array}{c}
                                     |0, \alpha\rangle \\
                                     0 \\
                                   \end{array}
                                 \right), \label{bosoniccoherentstate} \\
										{\cal D}(\alpha) |\Psi_1 \rangle = \left(
                                   \begin{array}{c}
																	0 \\
                                     D(\alpha)|0\rangle \\
                                   \end{array}
                                 \right) = \left(
                                   \begin{array}{c} 0 \\
                                     |0, \alpha\rangle \\
                                   \end{array}
                                 \right). \label{bosoniccoherentstate}						
\end{eqnarray}
The commutator
\begin{equation}
{\cal D}^\dagger(\alpha) A_0 {\cal D}(\alpha) = A_0 + \alpha I \hskip1cm\rightarrow \hskip1cm [A_0, {\cal D}(\alpha)] = \alpha {\cal D}(\alpha)
\end{equation}
applied to state $|\Psi\rangle$
\begin{equation}
A_0 ({\cal D}(\alpha) |\Psi\rangle) = \alpha ({\cal D}(\alpha) |\Psi\rangle) + {\cal D}(\alpha) A_0 |\Psi\rangle,
\end{equation}
gives the eigenvalue problem
\begin{equation}
A_0 ({\cal D}(\alpha) |\Psi\rangle) = \alpha ({\cal D}(\alpha) |\Psi\rangle),
\end{equation}
if the reference state $|\Psi\rangle$ is annihilated by operator $A_0$: $A_0 |\Psi \rangle =0$.
Therefore,
the coherent states, created from reference states $|\Psi_0 \rangle$ and $|\Psi_1\rangle$ and their superposition
$|\Psi\rangle = c_0 |\Psi_0\rangle + c_1 |\Psi_1\rangle
$
, satisfy
eigenvalue problem
\begin{equation}
A_0  (c_0|0\rangle_f + c_1 |1\rangle_f) \otimes |0,\alpha \rangle = \alpha \,  (c_0|0\rangle_f + c_1 |1\rangle_f) \otimes |0,\alpha \rangle ,
\label{A0eigenvalueproblem}
\end{equation}
and are separable. To create entangled super-coherent state we have to choose different reference state with entangled bosons and fermions.
In present work we consider the set of maximally entangled four Bell reference states (\ref{Lstates}), (\ref{Bstates}), as super-qubit states.

\begin{definition}
The Bell super-coherent states are defined as
\begin{eqnarray}
 |\alpha, L_\pm \rangle   \equiv     {\cal D}(\alpha) |L_\pm \rangle, \hskip0.5cm   |\alpha, B_\pm \rangle  \equiv   {\cal D}(\alpha) |B_\pm \rangle.
\end{eqnarray}
\end{definition}

\begin{prop}
The Bell super-coherent states  are eigenstates of corresponding supersymmetric annihilation operators
\begin{eqnarray}
A_1 |\alpha, L_-\rangle &=& \alpha |\alpha, L_-\rangle,\hskip0.5cm A_{-1} |\alpha, L_+\rangle = \alpha |\alpha, L_+\rangle,\\
A^T_1 |\alpha, B_-\rangle &=& \alpha |\alpha, B_-\rangle,\hskip0.5cm A^T_{-1} |\alpha, B_+\rangle = \alpha |\alpha, B_+\rangle.
\end{eqnarray}
The states are orthonormal and maximally entangled.
In explicit form the states are expressed as
\begin{eqnarray}
|\alpha,  L_\pm \rangle & =&\frac{1}{\sqrt{2}} (|0\rangle_f |1, \alpha\rangle \pm
|1\rangle_f |0, \alpha\rangle), \\
|\alpha, B_\pm \rangle &=&\frac{1}{\sqrt{2}} (|0\rangle_f |0, \alpha\rangle \pm
|1\rangle_f |1, \alpha\rangle),
\end{eqnarray}
in terms of the displaced Fock states
\begin{eqnarray}
|0, \alpha \rangle &=& D(\alpha) |0\rangle =e^{-\frac{1}{2} |\alpha|^2} |\alpha \rangle,\\
|1, \alpha \rangle &=& D(\alpha) |1\rangle=e^{-\frac{1}{2} |\alpha|^2} (\frac{d}{d\alpha} |\alpha \rangle - \bar\alpha |\alpha \rangle ).
\end{eqnarray}
Here $|\alpha \rangle$ is the Glauber coherent state (not normalized).
\end{prop}
The linear combination of these, maximally entangled  states with orthogonal separable states produces the set of four supercoherent states.
These states are created  by displacement operator, acting on super-qubit reference states.

\begin{prop}
The states (\ref{refLpm}) annihilated by  $A_{1}$ and $A_{-1}$ operators correspondingly, as  in (\ref{annihilateL}), determine the pair of
super-coherent states
\begin{equation}
           |\alpha, C, \phi\rangle_{L_\pm} \equiv      {\cal D}(\alpha) |0, C, \phi \rangle_{L_\pm},
\end{equation}
which are eigenstates of super annihilation operators
\begin{eqnarray}
A_1 |\alpha, C, \phi\rangle_{L_-} = \alpha |\alpha, C, \phi\rangle_{L_-}, \hskip0.5cm A_{-1} |\alpha, C, \phi\rangle_{L_+} = \alpha |\alpha, C, \phi\rangle_{L_+}
\end{eqnarray}
\end{prop}
\begin{prop}
The pair of reference states (\ref{refBpm}), annihilated by Eq. (\ref{annihilateB}), gives the pair of super-coherent states
\begin{equation}
     |\alpha, C, \phi\rangle_{B_\pm}  \equiv {\cal D}(\alpha) |0, C, \phi \rangle_{B_\pm},
\end{equation}
which are eigenstates of operators
\begin{eqnarray}
A^T_1 |\alpha, C, \phi\rangle_{B_-} = \alpha |\alpha, C, \phi\rangle_{B_-}, \hskip0.5cm A^T_{-1} |\alpha, C, \phi\rangle_{B_+} = \alpha |\alpha, C, \phi\rangle_{B_+}
\end{eqnarray}
\end{prop}
Then we have following definition.
\begin{definition}
The super-coherent states as displaced super-qubit states
\begin{eqnarray}
|\alpha, C, \phi \rangle_{L_{-}} = \sqrt{1-C} |0\rangle_f \otimes |0,\alpha\rangle + \sqrt{C} e^{i\phi} |\alpha, L_{-} \rangle, \label{Lminusalpha}\\
|\alpha, C, \phi \rangle_{L_{+}} = \sqrt{1-C} |0\rangle_f \otimes |0,\alpha\rangle + \sqrt{C} e^{i\phi} |\alpha, L_{+} \rangle, \label{Lplusalpha}\\
|\alpha, C, \phi \rangle_{B_{-}} = \sqrt{1-C} |1\rangle_f \otimes |0,\alpha\rangle + \sqrt{C} e^{i\phi} |\alpha, B_{-} \rangle, \label{Bminusaplha}\\
|\alpha, C, \phi \rangle_{B_{+}} = \sqrt{1-C} |1\rangle_f \otimes |0,\alpha\rangle + \sqrt{C} e^{i\phi} |\alpha, B_{+} \rangle. \label{Bplusalpha}
\end{eqnarray}
we call as the super-Bell based states.
\end{definition}
On the super-Bloch sphere
these states take form
\begin{eqnarray}
|\alpha, \theta, \phi \rangle_{L_{\mp}} &=& \cos \frac{\theta}{2} \left(
     \begin{array}{c}
      |0, \alpha \rangle \\
        0 \\
       \end{array}
      \right) + \sin \frac{\theta}{2} e^{i\phi} \frac{1}{\sqrt{2}}\left(
     \begin{array}{c}
      |1, \alpha \rangle \\
        \mp|0, \alpha \rangle  \\
       \end{array}
      \right) \label{supercoherentL1}, \\
|\alpha, \theta, \phi \rangle_{B_{\mp}} &=& \cos \frac{\theta}{2} \left(
     \begin{array}{c}
      0 \\
        |0, \alpha \rangle\\
       \end{array}
      \right) + \sin \frac{\theta}{2} e^{i\phi} \frac{1}{\sqrt{2}}\left(
     \begin{array}{c}
      |0, \alpha \rangle \\
        \mp|1, \alpha \rangle  \\
       \end{array}
      \right), \label{supercoherentB1}
\end{eqnarray}
or explicitly
\begin{eqnarray}
|\alpha, \theta, \phi \rangle_{L_{\mp}} &=& \cos \frac{\theta}{2} e^{-\frac{|\alpha|^2}{2}}\left(
     \begin{array}{c}
      |\alpha \rangle \\
        0 \\
       \end{array}
      \right) + \sin \frac{\theta}{2} e^{i\phi} \frac{e^{-\frac{|\alpha|^2}{2}}}{\sqrt{2}}\left(
     \begin{array}{c}
       |\alpha\rangle' - \bar\alpha|\alpha \rangle\\
        \mp|\alpha \rangle  \\
       \end{array}
      \right), \nonumber \\
			|\alpha, \theta, \phi \rangle_{B_{\mp}} &=& \cos \frac{\theta}{2} e^{-\frac{|\alpha|^2}{2}}\left(
     \begin{array}{c}
      0 \\
        |\alpha \rangle \\
       \end{array}
      \right) + \sin \frac{\theta}{2} e^{i\phi} \frac{e^{-\frac{|\alpha|^2}{2}}}{\sqrt{2}}\left(
     \begin{array}{c} |\alpha \rangle \\
        \mp |\alpha\rangle' \pm \bar\alpha|\alpha \rangle \\
       \end{array}
      \right). \nonumber
\end{eqnarray}
The states
are eigenstates of super-annihilation operators
\begin{eqnarray}
A_{\pm 1} |\alpha, \theta, \phi \rangle_{L_\mp} &=& \alpha |\alpha, \theta, \phi \rangle_{L_\mp}, \\
A^T_{\pm 1} |\alpha, \theta, \phi \rangle_{B_\mp} &=& \alpha |\alpha, \theta, \phi \rangle_{B_\mp},
\end{eqnarray}
with inner products
\begin{equation}
{_{L_{+}}}\langle \alpha, \theta, \phi | \alpha, \theta, \phi \rangle_{L_-} = \cos^2 \frac{\theta}{2} = {_{B_{+}}}\langle \alpha, \theta, \phi | \alpha, \theta, \phi \rangle_{B_-}.
\end{equation}
It is noted that supercoherent state $|\alpha, \theta, \phi \rangle_{L_{-}} $, for angle $\phi' = \phi + \pi $ coincides with the one, derived early in \cite{aragone}.

\section{Entanglement of Supercoherent States}

Here we are going to calculate entanglement of supercoherent states. As a first step, we calculate concurrence for the reference states (\ref{refLpm1}), (\ref{refBpm1}).
Then we show that concurrence is independent of action of the displacement operator on the states, and as follows it is independent of $\alpha$. As a result we find that the super-qubit reference state   and the corresponding super-coherent state have the same concurrence.

\subsection{Entanglement of Super-qubit States}

First we calculate entanglement of the superqubit states (\ref{refLpm2}).
  For these states
 \begin{equation}
|0, \theta, \phi \rangle_{L_{\pm}} = |0\rangle_f \otimes \left(\cos \frac{\theta}{2} |0 \rangle_b + \frac{1}{\sqrt{2}}\sin \frac{\theta}{2} e^{i\phi} |1\rangle_b \right)
\pm |1\rangle_f \otimes \frac{1}{\sqrt{2}}\sin \frac{\theta}{2} e^{i\phi} |0\rangle_b
\end{equation}
 the reduced density matrices are expressed by the same form, but in fermionic $|0\rangle_f, |1\rangle_f$ (two-component) and bosonic $|0\rangle_b, |1\rangle_b$ (infinite component) states,
\begin{eqnarray}
\rho_b = \rho_f =    \nonumber\\
 (\cos^2 \frac{\theta}{2} + \frac{1}{2} \sin^2 \frac{\theta}{2}) |0 \rangle \langle 0 | + \frac{1}{2} \sin^2 \frac{\theta}{2} |1 \rangle \langle 1 |
+ \frac{1}{2 \sqrt{2}} \sin \theta (e^{-i\phi} |0\rangle \langle 1 | +  e^{i\phi} |1\rangle \langle 0 |) \nonumber
\end{eqnarray}
so that
 \begin{equation}
tr \rho_b^2  = tr \rho_f^2 = 1 - \frac{1}{2} \sin^4 \frac{\theta}{2}.
\end{equation}

Comparing with (\ref{concurrence}) we find the concurrence for the reference states (\ref{refLpm2})
\begin{equation}
C = \sin^2 \frac{\theta}{2}. \label{refstateconcurrence}
\end{equation}

The result can be obtained also from general formula (\ref{genericconcurrence}) by identification with reference states (\ref{refLpm1}), written in terms of $C$,
\begin{equation}
|\psi_0 \rangle = \sqrt{1-C} |0\rangle + \frac{1}{\sqrt{2}}\sqrt{C} e^{i\phi} |1\rangle,\hskip1cm |\psi_1 \rangle =
\pm\frac{1}{\sqrt{2}} \sqrt{C} e^{i\phi} |0\rangle
\end{equation}
so that
\begin{eqnarray}
\langle\psi_0 | \psi_0 \rangle = 1 - \frac{1}{2} C, \hskip1cm \langle\psi_1 | \psi_1 \rangle = \frac{1}{2} C, \nonumber \\
\langle\psi_0 | \psi_1 \rangle = \overline{\langle\psi_1 | \psi_0 \rangle} = \pm\frac{1}{ \sqrt{2}} \sqrt{C (1-C)} e^{i\phi}. \nonumber
\end{eqnarray}
By calculating determinant (\ref{genericconcurrence}) we obtain formula (\ref{refstateconcurrence}).

	The formula implies that on the super-Bloch sphere the concurrence is monotonically increasing function of $\theta$, so that the minimal value $C =0$
at the north pole ($\theta =0$) corresponds to separable state $|\Psi_0\rangle$, while the maximally entangled state with $C =1$
relates to the south pole ($\theta = \pi$). On the equator ($\theta = \frac{\pi}{2}$) concurrence of the states  is equal $C =\frac{1}{2}$. Equation (\ref{refstateconcurrence})
justifies representation (\ref{refLpm1}) of reference states by concurrence $C$ and
shows that entanglement
is independent of angle $\phi$.

The same results for concurrence we obtain in case of the second couple of reference states (\ref{refBpm1}) or (\ref{refBpm2}). Thus, we have following proposition.

\begin{prop}
The concurrence $C$, $0 \le C \le 1$, for four reference states (\ref{refLpm2}) and (\ref{refBpm2}) is equal
\begin{equation}
C = \sin^2 \frac{\theta}{2}.
\end{equation}
The states can be parametrized by this concurrence as in (\ref{refLpm1}) and (\ref{refBpm1}).
\end{prop}

In the next section we show that the same formula for concurrence is valid for the supersymmetric coherent states (\ref{supercoherentL1}).

\subsection{Entanglement  for Displaced States}

An arbitrary normalized state $| \Phi \rangle$ from $H_f \otimes H_b$
\begin{equation}
|\Phi \rangle = \sum_{i=0}^1 \sum_{n=0}^\infty c_{i n} |i\rangle_f \otimes |n\rangle,
\end{equation}
where
\begin{equation}
\sum_{i=0}^1 \sum_{n=0}^\infty |c_{i n}|^2 =1,
\end{equation}
after application of the displacement operator ${\cal D}(\alpha)$ becomes
\begin{equation}
|\Phi, \alpha \rangle = {\cal D}(\alpha)|\Phi \rangle=\sum_{i=0}^1 \sum_{n=0}^\infty c_{i n} |i\rangle_f \otimes D(\alpha)|n\rangle =
\sum_{i=0}^1 \sum_{n=0}^\infty c_{i n} |i\rangle_f \otimes |n, \alpha\rangle,
\end{equation}
where $|n, \alpha\rangle = D(\alpha)|n\rangle$ are displaced Fock states.
This can be rewritten in two forms according to following propositions.

\begin{prop}
For an arbitrary state from $H_f \otimes H_b$, represented as
\begin{equation}
|\Phi \rangle = |0\rangle_f \otimes |\psi_0\rangle + |1\rangle_f \otimes |\psi_1\rangle
\end{equation}
by two states in Fock space
\begin{equation}
|\psi_0 \rangle = \sum^\infty_{n=0} c_{0 n} |n\rangle, \hskip1cm |\psi_1 \rangle = \sum^\infty_{n=0} c_{1 n} |n\rangle,
\end{equation}
the displaced state is
\begin{equation}
|\Phi, \alpha \rangle = {\cal D}(\alpha)|\Phi \rangle=|0\rangle_f \otimes |\psi_0, \alpha\rangle + |1\rangle_f \otimes |\psi_1, \alpha\rangle
\end{equation}
where
\begin{equation}
|\psi_0, \alpha \rangle = \sum^\infty_{n=0} c_{0 n} |n, \alpha\rangle, \hskip1cm |\psi_1, \alpha \rangle = \sum^\infty_{n=0} c_{1 n} |n, \alpha\rangle,
\end{equation}
and the displaced Fock states are
$|n, \alpha\rangle = D(\alpha) |n \rangle$.
The last states satisfy orthonormality conditions
\begin{equation}
\langle m, \alpha | n, \alpha \rangle = \langle m |D^\dagger(\alpha) D(\alpha) |n \rangle = \langle m| n\rangle = \delta_{m n}
\end{equation}
and completeness relation
\begin{equation}
\sum^\infty_{n=0} |n, \alpha \rangle \langle n, \alpha | = D(\alpha)\sum^\infty_{n=0} |n\rangle \langle n|D^\dagger(\alpha)
= D(\alpha) D^\dagger(\alpha) = I.
\end{equation}
\end{prop}

\begin{prop}
For arbitrary state from $H_f \otimes H_b$, represented by sum of infinite number of qubits
\begin{equation}
|\Phi \rangle = \sum_{n=0}^\infty  |\varphi_n \rangle \otimes |n \rangle
= \sum_{n=0}^\infty \left(
          \begin{array}{c}
            c_{0 n}  \\
            c_{1 n}\\
          \end{array}
        \right) \otimes |n \rangle
\end{equation}
the displaced state is
\begin{equation}
|\Phi \rangle = \sum_{n=0}^\infty  |\varphi_n \rangle \otimes |n, \alpha \rangle
= \sum_{n=0}^\infty \left(
          \begin{array}{c}
            c_{0 n}  \\
            c_{1 n}\\
          \end{array}
        \right) \otimes |n, \alpha \rangle.
\end{equation}

\end{prop}

As we have seen in (\ref{genericconcurrence}) the concurrence of a state depends on inner products of two bosonic states. By calculating the inner product for the displaced states
\begin{equation}
|\psi_i, \alpha \rangle = D(\alpha) |\psi_i \rangle, \hskip1cm i =0,1,
\end{equation}
we find that it is invariant under displacement operation and independent of $\alpha$,
\begin{equation}
\langle \psi_i, \alpha | \psi_j, \alpha \rangle = \langle \psi_i | D^\dagger (\alpha) D(\alpha)|\psi_j\rangle = \langle \psi_i | \psi_j, \rangle.
\end{equation}
This suggests that entanglement for generic state $|\Phi \rangle$ and the displaced one $|\Phi, \alpha \rangle = {\cal D}(\alpha) |\Phi \rangle$ is the same.
Indeed, from density matrix for displaced state
\begin{equation}
\rho(\alpha) = |\Phi, \alpha \rangle \langle \Phi, \alpha | = {\cal D}(\alpha)|\Phi \rangle \langle \Phi | {\cal D}^\dagger (\alpha) =
{\cal D}(\alpha) \rho {\cal D}^\dagger (\alpha)
\end{equation}
  we get reduced density matrix
\begin{equation}
\rho_b(\alpha) = | \psi_0, \alpha \rangle \langle\psi_0, \alpha | + | \psi_1, \alpha \rangle \langle\psi_1, \alpha | =
D(\alpha)(| \psi_0 \rangle \langle\psi_0 | + | \psi_1 \rangle \langle\psi_1 |) D^\dagger(\alpha), \nonumber
\end{equation}
so that
$\rho_b(\alpha) = D(\alpha) \rho_b D^\dagger(\alpha)$
and
$
\rho^2_b(\alpha) = D(\alpha) \rho^2_b D^\dagger(\alpha)
$.
By taking trace from both sides we find
$
tr \rho^2_b(\alpha) = tr \rho^2_b
$.
This shows that the concurrence
$
C^2 = 2 (1 - tr \rho^2_b)
$
and entanglement
for both states is the same and don't depends on complex parameter $\alpha$.
Therefore, we have following proposition.

\begin{prop}
The concurrences (entanglement) for state $|\Phi \rangle$ and the displaced state $|\Phi, \alpha \rangle = {\cal D}(\alpha) |\Phi \rangle$
are equal.
\end{prop}

\begin{cor}
For supersymmetric coherent states $|\alpha, C, \phi \rangle_{L_{\mp}}$,
$|\alpha, C, \phi \rangle_{B_{\mp}}$, defined in (\ref{Lminusalpha})-(\ref{Bplusalpha}),
 the concurrence is independent of $\alpha$ and is equal
\begin{equation}
C = \sin^2 \frac{\theta}{2}.
\end{equation}
For these states, the concurrence $C = p_1$ coincides with the probability of transition
to maximally entangled states and represents the geometric probability, as relative area of spherical cup on super-Bloch sphere $C = A_\theta/A$.
\end{cor}

In Figure 1 we show the concurrence $C$ and the von Neuman entropy $E$ as functions of angle $\theta$ on the super-Bloch sphere.

\begin{figure}[h]
  \centering
  \includegraphics[page=1, width= 0.5\textwidth]{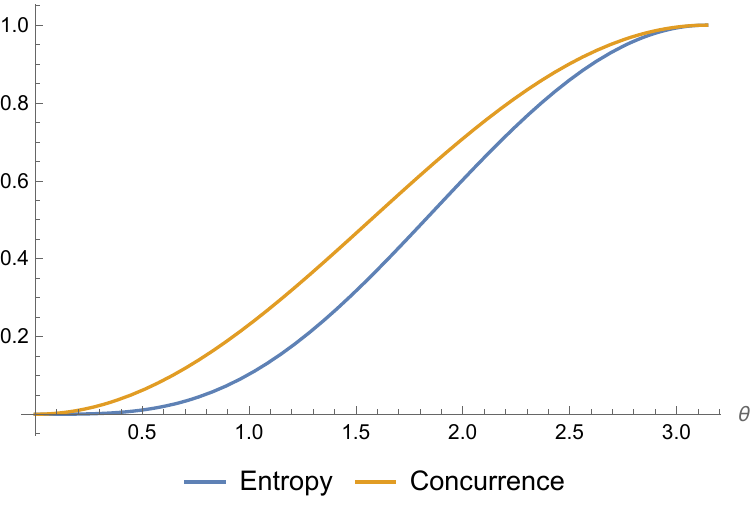}
  \caption{Concurrence and Entanglement versus angle $\theta$ on super-Bloch sphere}
  \label{fig:concurrence}
\end{figure}


As an example,  by using determinant formula (\ref{genericconcurrence}) for Hermitian metric, with two states
\begin{equation}
|\psi_0, \alpha \rangle = \sqrt{1-C} |0, \alpha\rangle + \frac{1}{\sqrt{2}}\sqrt{C} e^{i\phi} |1, \alpha\rangle,\hskip0.5cm |\psi_1, \alpha \rangle =
\pm\frac{1}{\sqrt{2}} \sqrt{C} e^{i\phi} |0, \alpha\rangle
\end{equation}
we find the same  concurrence for states (\ref{Lminusalpha}),  (\ref{Lplusalpha}),
 The second state is the  Glauber coherent state, while the first one is superposition of Glauber state with the one photon added coherent state. The last one is adding non-classical property to the coherent state, and as we can see it is responsible for entanglement between fermions and bosons in supercoherent state.

\subsection{Orthogonality of Super Coherent States}

Here we calculate the inner product of two super coherent states with the same position on the super Bloch sphere $(\theta, \phi)$
and show that in contrast to the Glauber coherent states, they can be orthogonal.
The product formulas for our displacement operators
\begin{eqnarray}
{\cal D}(\alpha) {\cal D}(\beta) = e^{2 i \Im (\alpha \bar\beta)} {\cal D}(\beta) {\cal D}(\alpha)
= e^{ i \Im (\alpha \bar\beta)} {\cal D}(\alpha + \beta)
\end{eqnarray}
give
\begin{eqnarray}
\langle \beta, \theta, \phi | \alpha, \theta, \phi \rangle &=& \langle 0, \theta, \phi | {\cal D}^\dagger (\beta) {\cal D}(\alpha) |0, \theta, \phi \rangle = \\
e^{-i \Im (\beta \bar\alpha)} \langle 0, \theta, \phi |  {\cal D}(\alpha -\beta) |0, \theta, \phi \rangle
&=& e^{-i \Im (\beta \bar\alpha)} \langle 0, \theta, \phi  |\alpha - \beta, \theta, \phi \rangle.
\end{eqnarray}
By using matrix elements
\begin{eqnarray}
\langle 0 |D(\alpha) |0 \rangle &=& e^{-\frac{1}{2}|\alpha|^2},\hskip1cm \langle 1 |D(\alpha) |0 \rangle = \alpha e^{-\frac{1}{2}|\alpha|^2},\\
\langle 0 |D(\alpha) |1 \rangle &=& -\bar\alpha e^{-\frac{1}{2}|\alpha|^2},\hskip1cm \langle 1 |D(\alpha) |1 \rangle = (1-|\alpha|^2)\alpha e^{-\frac{1}{2}|\alpha|^2},
\end{eqnarray}
we find
\begin{eqnarray}
{_{L_\pm}}\langle \beta, \theta, \phi | \alpha, \theta, \phi \rangle_{L_\pm} = \nonumber \\
e^{-i \Im (\beta \bar\alpha)} e^{-\frac{1}{2}|\alpha - \beta|^2} \left(
1 - \frac{\sin \theta}{2\sqrt{2}} ((\bar\alpha - \bar\beta) e^{i\phi} - (\alpha -\beta) e^{-i\phi}) - \frac{|\alpha-\beta|^2}{2}\sin^2 \frac{\theta}{2}\right). \nonumber
\end{eqnarray}
In the limiting case $\theta =0$ (separable state at the north pole) we have the usual inner product formula for bosonic coherent states
\begin{equation}
\langle \beta, 0, \phi | \alpha, 0, \phi \rangle = e^{-\frac{1}{2}|\alpha|^2} e^{-\frac{1}{2}|\beta|^2} e^{\bar\beta \alpha},
\end{equation}
which is never  zero. In another limit $\theta = \pi$ (maximally entangled state at the south pole) we have

\begin{equation}
{_{L_\pm}}\langle \beta, \pi, \phi | \alpha, \pi, \phi \rangle_{L_\pm} = \left(1 - \frac{1}{2}{|\alpha - \beta|^2}\right) e^{-\frac{1}{2}|\alpha|^2} e^{-\frac{1}{2}|\beta|^2} e^{\bar\beta \alpha}.
\end{equation}

In contrast with pure bosonic coherent states, in this case  the states can be orthogonal. The  set of orthogonal maximally entangled states satisfies
condition
\begin{equation}
|\alpha - \beta|^2 =2,
\end{equation}
and belongs to the circle in complex plane with radius $r = \sqrt{2}$ around point $\alpha$. Then, every state on the circle,  parametrized
by $\beta = \alpha + \sqrt{2} \,e^{it}$, $0 \le t \le 2\pi$ is orthogonal to state $\alpha$. From this set it is always possible to choose the pair of states
$\beta_1$ and $\beta_2$ at distance $|\beta_1 - \beta_2| =\sqrt{2}$ and as a result, orthogonal to each other. So,  we have three mutually orthogonal states $\alpha$, $\beta_1 = \alpha + \sqrt{2} e^{i t_1}$
and $\beta_2 = \alpha + \sqrt{2} e^{i (t_1 + \frac{\pi}{3})}$, located at vertices of equilateral triangle.

In general case of arbitrary states $\alpha$ and $\beta$, the orthogonality condition takes the complex form
\begin{equation}
\frac{1}{2}  |w|^2 \sin^2 \frac{\theta}{2}  + \frac{1}{2\sqrt{2}}  (w - \bar w)\sin \theta  -1 =0,
\end{equation}
where
$
w \equiv (\alpha - \beta) e^{-i\phi}
$,
equivalent to the pair of real equations
\begin{eqnarray}
(w - \bar w)\sin \theta &=& 0, \\
\frac{1}{2}  |w|^2 \sin^2 \frac{\theta}{2} &=& 1.
\end{eqnarray}
It has solutions for $\theta = \pi$, considered above. In addition, for arbitrary
$0 < \theta < \pi$, such that $\sin \theta \neq 0$, and $w = \bar w$ is real,
we have two solutions
\begin{equation}
w_{1,2} = \pm \frac{\sqrt{2}}{\sin \frac{\theta}{2}},
\end{equation}
giving in terms of concurrence $C$,
\begin{equation}
\alpha - \beta = \pm \sqrt{\frac{2}{C}} \, e^{i\phi}.
\end{equation}
This implies that for any state $\alpha$ exists two (antipodal) states $\beta_+$ and $\beta_-$, orthogonal to the state $\alpha$,
\begin{equation}
\beta_+ = \alpha + \sqrt{\frac{2}{C}} \,e^{i\phi}, \hskip1cm \beta_- = \alpha - \sqrt{\frac{2}{C}} \,e^{i\phi}.
\end{equation}
These states exist for any level of entanglement $0 < C < 1$ and for separable states with $C =0$ they move to infinity. For maximally entangled states
with $C =1$, in addition to this pair, appears the circle of states, orthogonal to state $\alpha$. This result relates entanglement of super-coherent states with orthogonality,
so that to be orthogonal, the states should be necessarily entangled and non-classical.

Similar calculations for second pair of states $|\alpha, \theta, \phi \rangle_{B_\pm}$ give the same conditions of orthogonality.

\section{Time Independence of Entanglement}

Time dependence of coherent states is determined by evolution operator
\begin{equation}
{\cal U}(t) = e^{-i\omega H t} = \left(
 \begin{array}{cc}
  e^{-i\omega t N} &     0         \\ \\
    0 &  e^{-i\omega t (N+1)} \\
	\end{array} \right)    =  \left(
 \begin{array}{cc}
  1 &     0         \\ \\
    0 &  e^{-i\omega t} \\
	\end{array} \right)  \otimes e^{-i\omega t N}.
\end{equation}

\begin{prop}
The concurrence for arbitrary time dependent state $|\Phi(t) \rangle = {\cal U}(t) |\Phi \rangle$ is independent of time $C (t) = C$.
\end{prop}

\begin{prf}
For an arbitrary state  (\ref{BFstate}), decomposed as
\begin{equation}
|\Phi \rangle = |0\rangle_f \otimes |\psi_0\rangle + |1\rangle_f \otimes |\psi_1\rangle,
\end{equation}
where
\begin{equation}
|\psi_0 \rangle = \sum^\infty_{n=0} c_{0 n} |n\rangle,\,\,\,|\psi_1 \rangle = \sum^\infty_{n=0} c_{1 n} |n\rangle,
\end{equation}
the time dependent state is
\begin{equation}
|\Phi(t) \rangle = {\cal U}(t)|\Phi\rangle = |0\rangle_f \otimes |\psi_0(t)\rangle + e^{-i\omega t}|1\rangle_f \otimes |\psi_1(t)\rangle,
\end{equation}
 where $(a=0,1)$,
\begin{equation}
|\psi_a(t) \rangle = e^{-i\omega t N} |\psi_a\rangle = \sum^\infty_{n=0} c_{a n}(t) |n\rangle = \sum^\infty_{n=0} c_{a n} e^{-i\omega t n} |n\rangle.
\end{equation}
By calculating the inner products of these time dependent states and using  (\ref{genericconcurrence}) we have time independence of the
concurrence
\begin{eqnarray}
C(t) &=& 2 \sqrt{\left| det \left(
 \begin{array}{cc}
  \langle\psi_0 (t)| \psi_0(t) \rangle &        \langle\psi_0 (t) | \psi_1 (t) \rangle   e^{-i\omega t}    \\ \\
     \langle\psi_1 (t)| \psi_0 (t)\rangle e^{i\omega t} &  \langle\psi_1 (t)| \psi_1 (t)\rangle  \\
	\end{array} \right) \,\right|} \\
	&=& 2 \sqrt{ \left| det \left(
 \begin{array}{cc}
  \langle\psi_0 | \psi_0 \rangle &        \langle\psi_0  | \psi_1  \rangle   e^{-i\omega t}    \\ \\
     \langle\psi_1 | \psi_0 \rangle e^{i\omega t} &  \langle\psi_1 | \psi_1 \rangle  \\
	\end{array} \right) \,\right|} = C
\end{eqnarray}
\end{prf}

\section{Uncertainity Relations and Entanglement  on Super-Bloch Sphere}

Here we calculate the uncertainty relations for quartet of supercoherent states $|\alpha, \theta, \phi \rangle_{L_\pm}$
$|\alpha, \theta, \phi \rangle_{B_\pm}$. Calculations of averages for states $|\alpha, \theta, \phi \rangle_{L_\pm}$ and
$|\alpha, \theta, \phi \rangle_{B_+}$ give the same results, this is why we suppress the index of the states. The sign difference appearing for state  $|\alpha, \theta, \phi \rangle_{B_-}$ would be noticed in proper place.
The coordinate and momentum operators we define as
$
X =  I_f \otimes \frac{1}{\sqrt{2}}    (a + a^\dagger)$,
$ P =  I_f \otimes \frac{i}{\sqrt{2}}   (a^\dagger - a)$,
transformed by displacement operator (\ref{superdisplacement}) to
\begin{eqnarray}
{\cal D}^\dagger(\alpha) X {\cal D}(\alpha) = X + I_f \otimes \frac{\alpha + \bar\alpha}{\sqrt{2}} = X + I_f \otimes \sqrt{2} \Re \alpha, \\
{\cal D}^\dagger(\alpha) P {\cal D}(\alpha) = X + I_f \otimes i\frac{\bar\alpha - \alpha}{\sqrt{2}} = P + I_f \otimes \sqrt{2} \Im \alpha.
\end{eqnarray}
The mean values of the operators in supercoherent state
\begin{equation}
|\alpha, C, \phi \rangle = \sqrt{1 -C} |\alpha, \Psi_0\rangle + \sqrt{C} e^{i\phi} |\alpha, L_\pm\rangle
\end{equation}
 reduce to the forms
\begin{eqnarray}
\langle \alpha, C, \phi | X | \alpha, C, \phi \rangle &=& \langle 0, C, \phi | {\cal D}^\dagger(\alpha) X {\cal D}(\alpha) | 0, C, \phi \rangle=\sqrt{2} \Re \alpha + \langle 0, C, \phi | X | 0, C, \phi \rangle, \nonumber\\
\langle \alpha, C, \phi | P | \alpha, C, \phi \rangle &=& \langle 0, C, \phi | {\cal D}^\dagger(\alpha) P {\cal D}(\alpha) | 0, C, \phi \rangle=\sqrt{2} \Im \alpha + \langle 0, C, \phi | P | 0, C, \phi \rangle, \nonumber
\end{eqnarray}
which include the mean values in the reference super-qubit state,
\begin{equation}
|0, C, \phi \rangle = \sqrt{1 -C} \left(
          \begin{array}{c}
            |0\rangle  \\
            0\\
          \end{array}
        \right) + \sqrt{C} e^{i\phi} \frac{1}{\sqrt{2}}
				\left(
          \begin{array}{c}
            |1\rangle  \\
            \pm|0 \rangle\\
          \end{array}
        \right).
\end{equation}
For the last ones we get
\begin{eqnarray}
\langle 0, C, \phi | X | 0, C, \phi \rangle &=&  \sqrt{C (1-C)} \cos \phi, \label{Xaverage0}\\
\langle 0, C, \phi | P | 0, C, \phi \rangle &=& \sqrt{C (1 -C)} \sin \phi, \label{Paverage0}
\end{eqnarray}
valid for the first three states and including sign minus in the r.h.s. for the state $|\alpha, \theta, \phi \rangle_{B_-}$.
Then, we have
\begin{eqnarray}
\langle \alpha, C, \phi | X | \alpha, C, \phi \rangle &=& \sqrt{2} \Re \alpha +  \sqrt{C (1-C)} \cos \phi, \label{Xaverage}\\
\langle \alpha, C, \phi | P | \alpha, C, \phi \rangle &=& \sqrt{2} \Im \alpha +  \sqrt{C (1 -C)} \sin \phi. \label{Paverage}
\end{eqnarray}
In Figure 2 we plot average $\bar X = \langle \alpha, C, \phi | X | \alpha, C, \phi \rangle$  and in Figure 3,  $\bar P = \langle \alpha, C, \phi | P| \alpha, C, \phi \rangle$  as functions of the concurrence $C$ and angle $\phi$, where $\alpha = (1+i)/\sqrt{2}$.

\begin{figure}[h]
\begin{minipage}[t]{0.5\textwidth}
\centering
\vspace{0.5pt}
\includegraphics[width=\linewidth]{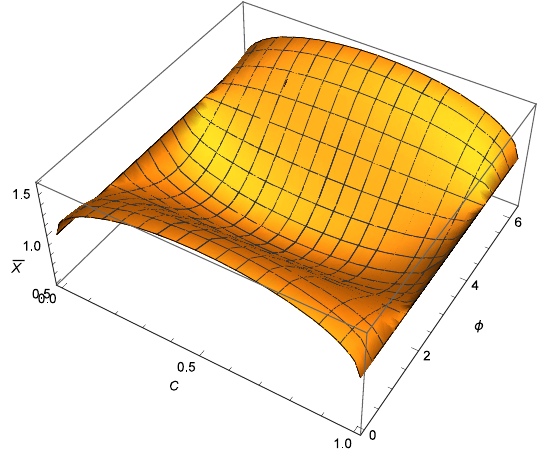}
\end{minipage}%
\hspace{0.05\textwidth}%
\begin{minipage}[t]{0.4\textwidth}
\centering
\vspace{0pt}
\includegraphics[width=\linewidth]{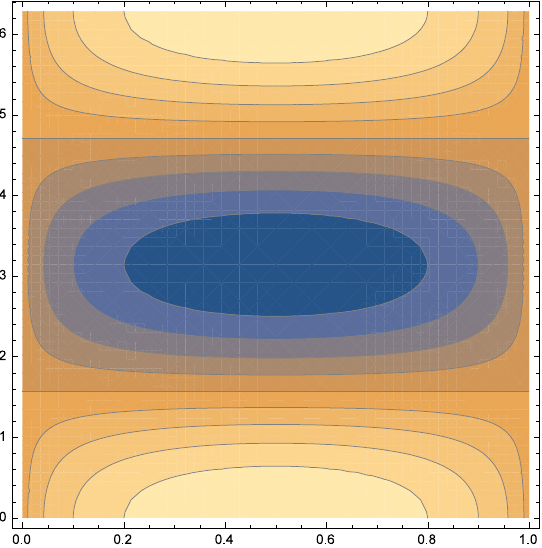}
\end{minipage}
\caption{The average value $\bar X$ as function of $C$ and $\phi$,  for $\alpha = (1+i)/\sqrt{2}$ : a) 3D plot b) Contour Plot}
\label{fig:your_common_figure}
\end{figure}

\begin{figure}[h]
\begin{minipage}[t]{0.5\textwidth}
\centering
\vspace{0.5pt}
\includegraphics[width=\linewidth]{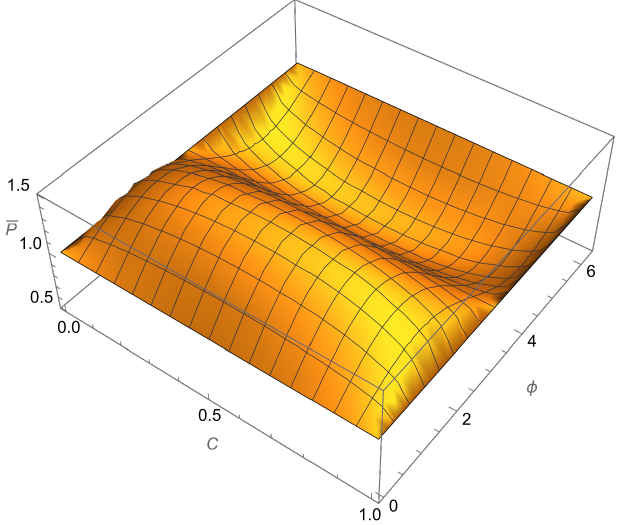}
\end{minipage}%
\hspace{0.05\textwidth}%
\begin{minipage}[t]{0.4\textwidth}
\centering
\vspace{0pt}
\includegraphics[width=\linewidth]{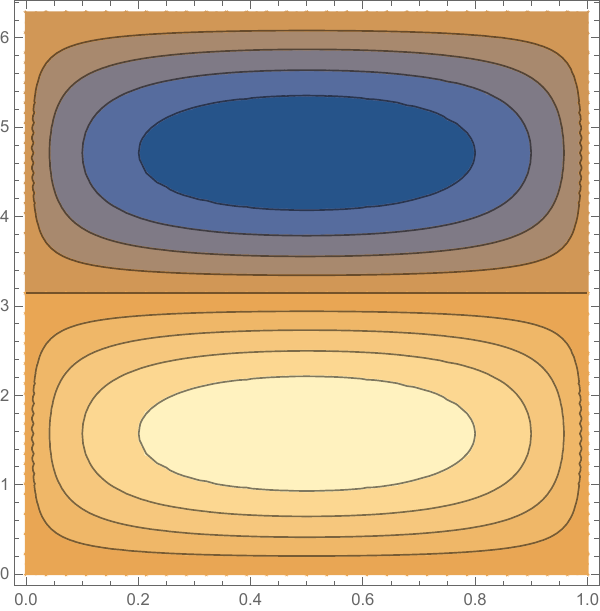}
\end{minipage}
\caption{The average value $\bar P$ as function of $C$ and $\phi$  for $\alpha = (1+i)/\sqrt{2}$: a) 3D plot b) Contour Plot}
\label{fig:your_common_figure}
\end{figure}

The averages of $a$ and $a^\dagger$  operators are
\begin{eqnarray}
\langle 0, C, \phi | I_f \otimes a | 0, C, \phi \rangle &=& \sqrt{\frac{C (1-C)}{2}} \,e^{i\phi}, \\
\langle 0, C, \phi | I_f \otimes a^\dagger | 0, C, \phi \rangle &=& \sqrt{\frac{C (1-C)}{2}}\, e^{-i\phi}.
\end{eqnarray}
and
\begin{eqnarray}
\langle \alpha, C, \phi | I_f \otimes a | \alpha, C, \phi \rangle &=& \alpha + \sqrt{\frac{C (1-C)}{2}} e^{i\phi}, \\
\langle \alpha, C, \phi | I_f \otimes a^\dagger | \alpha, C, \phi \rangle &=& \bar\alpha + \sqrt{\frac{C (1-C)}{2}} e^{-i\phi}.
\end{eqnarray}

For the states with vanishing average values
\begin{eqnarray}
\langle \alpha, C, \phi | X | \alpha, C, \phi \rangle =0, \hskip1cm
\langle \alpha, C, \phi | P | \alpha, C, \phi \rangle =0
\end{eqnarray}
this gives
\begin{equation}
\alpha = -\sqrt{\frac{C (1-C)}{2}} e^{i\phi}
\end{equation}
so that $|\alpha|^2 = C(1-C)/2$.

Equations  (\ref{Xaverage}), (\ref{Paverage}) show that for states with  $C =0$ and $C = 1$, the average position and momentum are the same as for the bosonic coherent states. It deviates  from classical averages, when  $0 < C < 1$ and the difference reaches maximal value for
$C = \frac{1}{2}$, corresponding to states on the equator of the super-Bloch sphere
\begin{equation}
|\alpha, \frac{1}{2}, \phi \rangle = \frac{1}{\sqrt{2}} ( |\alpha, \Psi_0\rangle +  e^{i\phi} |\alpha, L_\pm\rangle ). \label{maxrandom}
\end{equation}

To calculate the average of $X^2$ and $P^2$, we use
\begin{eqnarray}
\langle \alpha, C, \phi | I_f \otimes a^2 | \alpha, C, \phi \rangle &=&
\langle 0, C, \phi | {\cal D}^\dagger(\alpha) I_f \otimes a^2 {\cal D}(\alpha)| 0, C, \phi \rangle \\&=&
\langle 0, C, \phi |I_f \otimes (a + \alpha)^2 | 0, C, \phi \rangle,  \nonumber\\
\langle \alpha, C, \phi | I_f \otimes {a^\dagger}^2 | \alpha, C, \phi \rangle &=& \langle 0, C, \phi | {\cal D}^\dagger(\alpha) I_f \otimes {a^\dagger}^2 {\cal D}(\alpha)| 0, C, \phi \rangle \\&=& \langle 0, C, \phi |I_f \otimes (a^\dagger + \bar\alpha)^2 | 0, C, \phi \rangle, \nonumber
\end{eqnarray}
and
\begin{eqnarray}
\langle 0, C, \phi | I_f \otimes a^\dagger a | 0, C, \phi \rangle &=& \frac{1}{2} C, \nonumber\\
\langle 0, C, \phi | I_f \otimes a a^\dagger | 0, C, \phi \rangle &=& \frac{1}{2} C + 1, \nonumber
\end{eqnarray}
so that
\begin{eqnarray}
\langle \alpha, C, \phi | X^2 | \alpha, C, \phi \rangle =    \nonumber  \\
\frac{1}{2} [(\alpha + \bar\alpha)^2 + 2 \sqrt{2} (\alpha + \bar\alpha) \sqrt{C (1-C)} \cos \phi + 1 + C], \\
\langle \alpha, C, \phi | P^2 | \alpha, C, \phi \rangle =  \nonumber \\ \frac{1}{2} [-(\alpha - \bar\alpha)^2 - 2 \sqrt{2} i (\alpha - \bar\alpha) \sqrt{C (1-C)} \sin \phi + 1 + C] .
\end{eqnarray}

Calculating dispersions we have following Theorem.

\begin{thm} Dispersions of coordinate $X$ and momentum $P$ in super-coherent states $|\alpha, C, \phi\rangle_{L_\pm}$ and $|\alpha, C, \phi\rangle_{B_\pm}$
are the same and equal
\begin{eqnarray}
 \left( \Delta {X} \right)^{2} _{\alpha} \equiv \langle {X}^{2}  \rangle _{\alpha}-\langle {X} \rangle _{\alpha}^2 &=&  \frac{1}{2}(1 + C) - C (1-C) \cos^2 \phi , \label{X2}\\
 \left( \Delta {P} \right)^{2} _{\alpha}\equiv \langle {P}^{2}\rangle_{\alpha}-\langle {P} \rangle _{\alpha}^2 &=& \frac{1}{2} (1 + C) - C (1 -C) \sin^2 \phi \label{P2}.
\end{eqnarray}
They not depend of $\alpha$,
$\left( \Delta X \right)^2_\alpha = \left( \Delta X \right)^2_0$, and
$\left( \Delta P \right)^2_\alpha = \left( \Delta P \right)^2_0$.
\end{thm}

The dispersions  satisfy "the Pythagoras theorem" in the phase plane
\begin{equation}
\left( \Delta X \right)^2 + \left( \Delta P \right)^2 = 1 + C^2
\end{equation}
for the right triangle with sides, $\Delta X$, $\Delta P$ and hypotenuse $\sqrt{1 + C^2}$. Then, the uncertainty relation $\Delta X \Delta P = A$ is given by the area
of rectangle with diagonal $\sqrt{1 + C^2}$.
The sides of the triangle are bounded between
\begin{eqnarray}
\frac{1}{2} (1 -C + 2 C^2) \le (\Delta X)^2 \le \frac{1}{2} (1 + C),\\
\frac{1}{2} (1 -C + 2 C^2) \le (\Delta P)^2 \le \frac{1}{2} (1 + C).
\end{eqnarray}

The uncertainty relation for the supersymmetric coherent states are found as monotonically growing function of $C$,
 \begin{equation}
\Delta X \Delta P = \frac{1}{2} \sqrt{1 + C^2 + 2 C^3 + C^2 (1 - C)^2 \sin^2 2\phi}, \label{XP0}
\end{equation}
with small periodic dependence on angle $\phi$.
It is shown in Figure 4.

\begin{figure}[h]
\begin{minipage}[t]{0.5\textwidth}
\centering
\vspace{0.5pt}
\includegraphics[width=\linewidth]{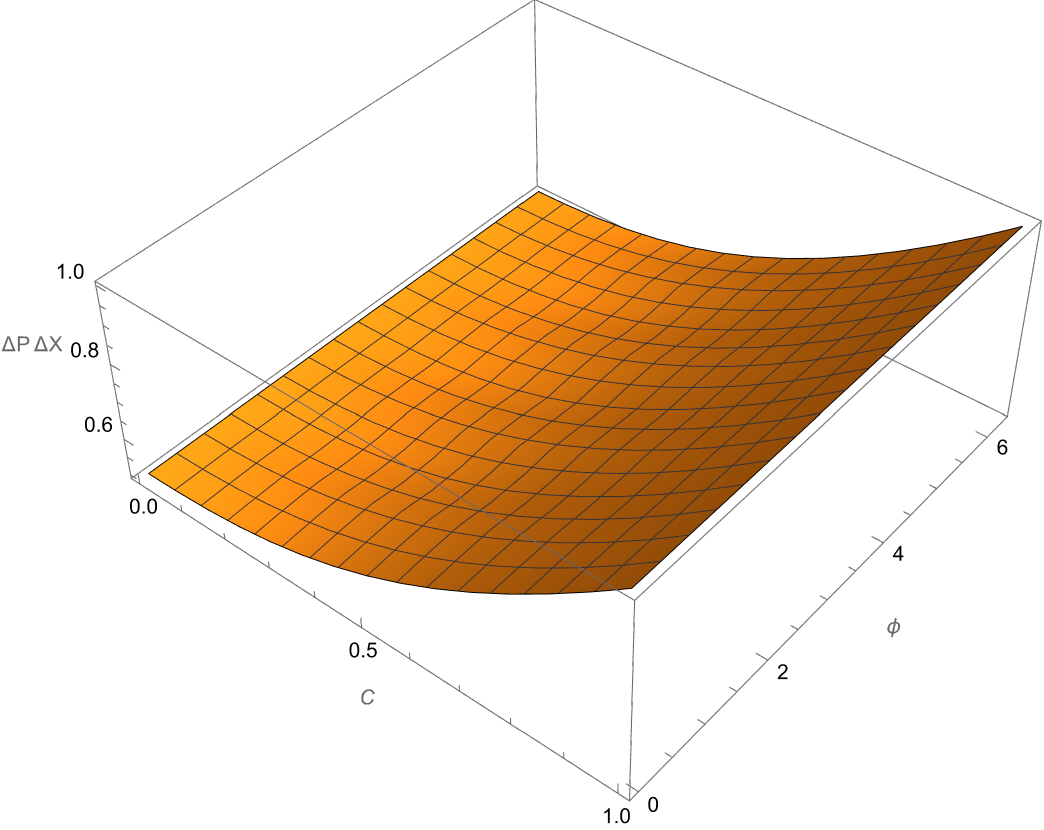}
\end{minipage}%
\hspace{0.05\textwidth}%
\begin{minipage}[t]{0.4\textwidth}
\centering
\vspace{0pt}
\includegraphics[width=\linewidth]{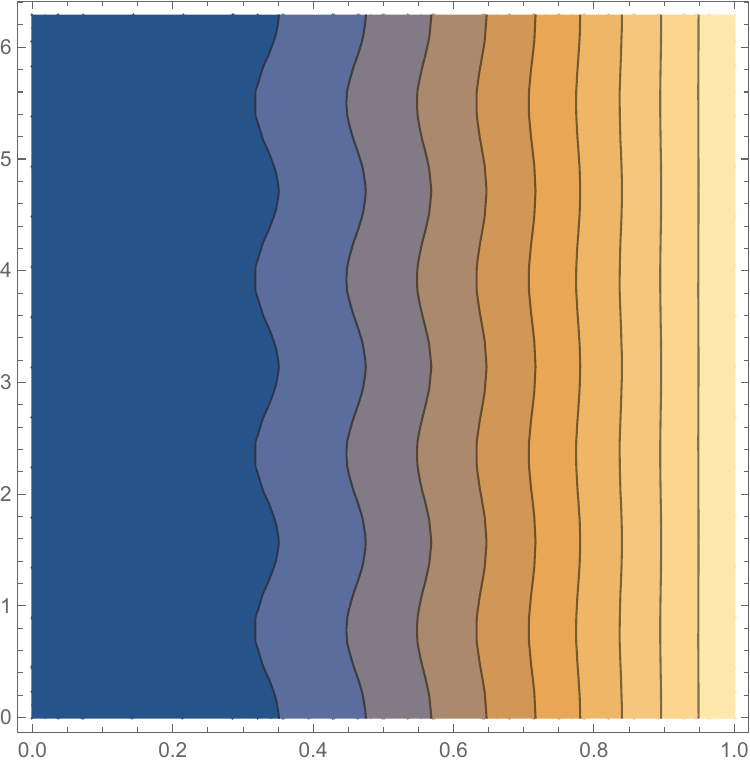}
\end{minipage}
\caption{Uncertainty relation   versus concurrence $C$ and angle $\phi$  : a) 3D plot b) Contour Plot}
\label{fig:your_common_figure}
\end{figure}

This implies inequality
\begin{equation}
\frac{1}{2} \sqrt{1 + C^2 + 2 C^3}\le \Delta X \Delta P \le \frac{1}{2} (1 + C^2),
\end{equation}
with minimum value at $\phi =0$ and the maximal one at $\phi = \frac{\pi}{4}, \frac{3\pi}{4}, \frac{5\pi}{4},\frac{7\pi}{4}$.
In the last case,
the triangle becomes isosceles triangle,
so that
dispersions are equal,
\begin{equation}
\left( \Delta X \right)^2 = \left( \Delta P \right)^2 = \frac{1 + C^2}{2}
\end{equation}
or
\begin{equation}
\Delta X  = \Delta P = \sqrt{\frac{1 + C^2}{2}}.
\end{equation}
The area of the square as the double area of the triangle  is maximal for fixed $C$ and gives uncertainty relation
\begin{equation}
\Delta X \Delta P = \frac{1+C^2}{2}. \label{square}
\end{equation}
It is noted from (\ref{X2}) and (\ref{P2}), that similarly to the bosonic coherent states \cite{klauder},
the dispersions are not dependent on $\alpha$, but on the reference super-qubit state, corresponding to $\alpha =0$,
(which is not the vacuum state), so that $\left( \Delta X \right)^2_\alpha = \left( \Delta X \right)^2_0$, and
$\left( \Delta P \right)^2_\alpha = \left( \Delta P \right)^2_0$.

The right hand side of equation (\ref{XP0}) is monotonically growing function of $C$,
bounded between $\frac{1}{2}$ and $1$ ,
\begin{equation}
\frac{1}{2} \leq \left( \Delta X \right)  \left( \Delta P \right)   \leq 1.
\end{equation}
The lower limit
\begin{eqnarray}
\label{uncertainityforAb}
 \left( \Delta X  \right)\left( \Delta P  \right) = \frac{1}{2}
\end{eqnarray}
corresponds to $C = 0$ and the state $|\alpha, \Psi_0\rangle$,
while the upper limit for $C = 1,$
\begin{equation}\label{uncertainityforAs}
 \left( \Delta X  \right)\left( \Delta P  \right)  = 1,
\end{equation}
 to the state $|\alpha, L_\pm\rangle$.
Obtained  relations show that for zero fermionic state the uncertainty reaches the minimal value, corresponding to pure bosonic coherent state as most classical quantum state and it is separable state with $C =0$.
Then,  mixing bosonic and fermionic degrees, due to nonclassical nature of fermions, increases non-classicality of the states and corresponding uncertainty.
It reaches maximal value for $C = 1$, which corresponds to maximally entangled bosonic and fermionic states as maximally non-classical states.

\subsection{Quadratic Squeezing of Coordinate and Momentum Uncertainties  }

As we have seen from uncertainty relation (\ref{XP0}), the product $(\Delta X \Delta P)^2$ reaches minimal value $\frac{1}{4}$ for $C =0$. This suggests that minimal uncertainty as in pure bosonic case of Glauber coherent states, should corresponds to   $(\Delta X)^2 = (\Delta P)^2 =\frac{1}{2}$.
But, it is not the case. In fact, depending on value of $\phi$, and $C$,  $(\Delta X)^2$ reaches local minima, smaller than $\frac{1}{2}$.
The uncertainty in $X$ as functions of two variables
\begin{eqnarray}
 \left( \Delta {X} \right)^{2}(C, \phi) \equiv f(C, \phi) &=&  \frac{1}{2}(1 + C) - C (1-C) \cos^2 \phi,
\end{eqnarray}
describes two dimensional surface. It is shown in Figure 5.
For this surface
we have conditions for first derivatives
\begin{eqnarray}
f_C(C, \phi) & =& \frac{1}{2} + (2 C -1) \cos^2 \phi =0, \nonumber \\
f_\phi(C, \phi) &=& C (1 -C) \sin 2\phi =0, \nonumber
\end{eqnarray}
giving two critical points
\begin{equation}
\phi = 0, \pi,\hskip1cm C = \frac{1}{4}, \label{critical1}
\end{equation}
and by using second derivatives
\begin{eqnarray}
f_{CC} = 2 \cos^2 \phi,\,\,\, f_{\phi \phi} = 2 C (1-C) \cos 2\phi, \,\,\, f_{C\phi} =f_{\phi C} = (1-2C) \sin 2\phi, \nonumber
\end{eqnarray}
we calculate the Gaussian curvature as determinant of the Hessian
\begin{equation}
H = f_{CC} f_{\phi\phi} - f^2_{C\phi} = 4 C(1-C) \cos2\phi \cos^2\phi - (1-2C)^2  \sin^2 2\phi.\nonumber
\end{equation}
For the critical points (\ref{critical1}) we get positive Gaussian curvature $H = \frac{3}{4}$ and due to $f_{CC} = 2 >0$, the local minimum.
The value of dispersions at these critical points is
\begin{equation}
\left( \Delta {X} \right)^{2} = f(\frac{1}{4}, 0) = f(\frac{1}{4}, \pi) = \frac{7}{16} < \frac{8}{16} = \frac{1}{2}, \hskip0.5cm \left( \Delta {P} \right)^{2} = \frac{5}{8} > \frac{4}{8} = \frac{1}{2}.
\label{criticaldispersion}
\end{equation}
The inequalities show that in the super-coherent states
\begin{equation}
 | \alpha, \frac{1}{4}, 0 \rangle_{L_\pm} = \frac{\sqrt{3}}{2} | \alpha, \Psi_0 \rangle + \frac{1}{2} |\alpha, L_\pm \rangle, \hskip0.5cm
| \alpha, \frac{1}{4}, \pi \rangle_{L_\pm} = \frac{\sqrt{3}}{2} | \alpha, \Psi_0 \rangle - \frac{1}{2} |\alpha, L_\pm \rangle
\end{equation}
the $X$ dispersion is maximally squeezed to value $\left( \Delta {X} \right)^{2} = \frac{7}{16} < \frac{1}{2}$, while $\left( \Delta {P} \right)^{2} = \frac{5}{8} > \frac{1}{2}$.
Positions of these states on the super-Bloch sphere are $(\theta =\frac{\pi}{3}, \phi =0)$ and $(\theta =\frac{\pi}{3}, \phi =\pi)$ correspondingly, and in complex plane representation at $z = \pm\frac{1}{\sqrt{3}}$.
\begin{figure}[h]
\begin{minipage}[t]{0.5\textwidth}
\centering
\vspace{0.5pt}
\includegraphics[width=\linewidth]{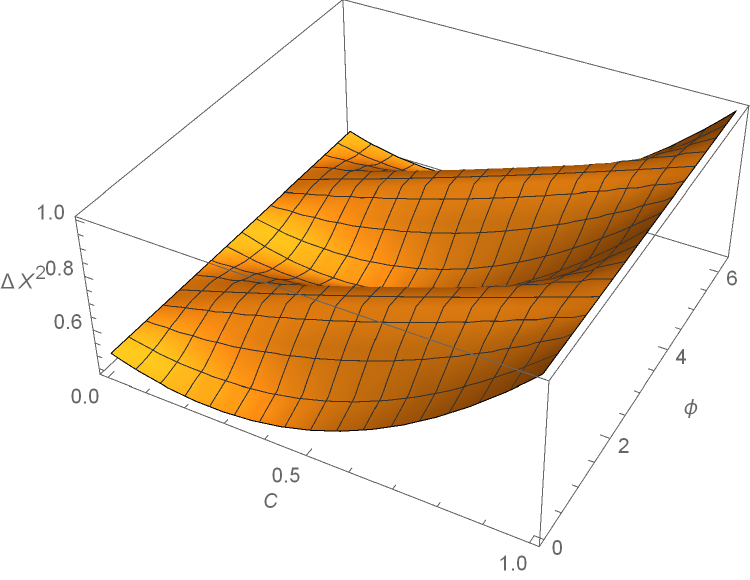}
\end{minipage}%
\hspace{0.05\textwidth}%
\begin{minipage}[t]{0.4\textwidth}
\centering
\vspace{0pt}
\includegraphics[width=\linewidth]{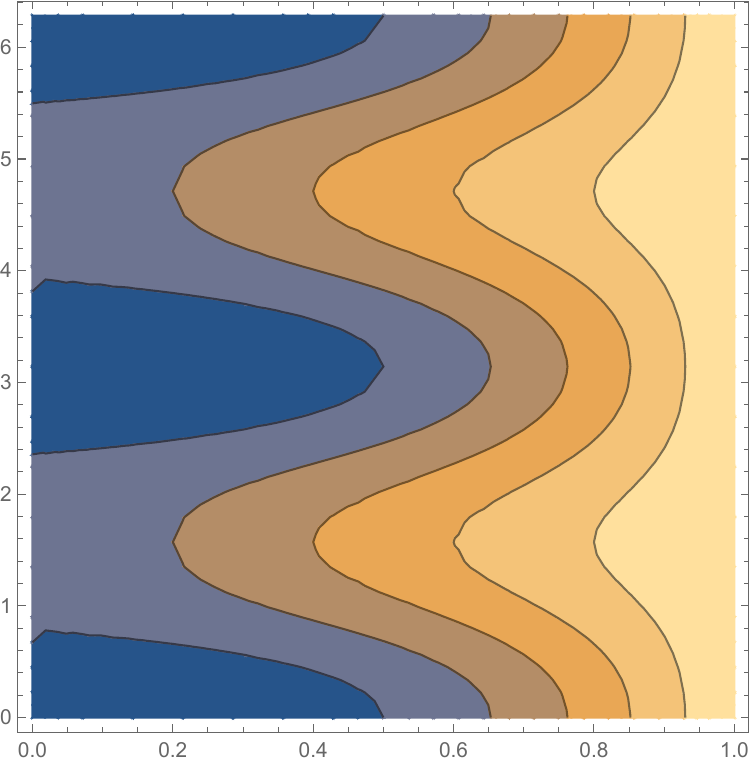}
\end{minipage}
\caption{Quadrature squeezing for dispersion $\Delta X^2$   versus concurrence $C$ and angle $\phi$  : a) 3D plot b) Contour Plot}
\label{fig:your_common_figure}
\end{figure}
In Figure 6, we plot dispersions versus the concurrence $C$ at angles $\phi =0$ and $\phi =\pi$.
Similar calculations for states
\begin{equation}
 | \alpha, \frac{1}{4}, \frac{\pi}{2} \rangle_{L_\pm} = \frac{\sqrt{3}}{2} | \alpha, \Psi_0 \rangle + \frac{i}{2} |\alpha, L_\pm \rangle, \hskip0.5cm
 | \alpha, \frac{1}{4}, \frac{3\pi}{2} \rangle_{L_\pm} = \frac{\sqrt{3}}{2} | \alpha, \Psi_0 \rangle - \frac{i}{2} |\alpha, L_\pm \rangle,
\end{equation}
at critical points on super-Bloch sphere $(\theta = \frac{\pi}{3}, \phi = \frac{\pi}{2})$, $(\theta = \frac{\pi}{3}, \phi = \frac{3\pi}{2})$ or in complex plane $z = \pm \frac{i}{\sqrt{3}}$,
give maximal squeezing for the momentum dispersion, $\left( \Delta {P} \right)^{2} = \frac{7}{16} < \frac{1}{2}$, $\left( \Delta {X} \right)^{2} = \frac{5}{8} > \frac{1}{2}$.
This quadrature squeezing is known for photon added coherent states, as non-classical property, and now we have established it also for boson-fermion entangled super-coherent states.

\begin{center}
\begin{figure}[h]
  \centering
  \includegraphics[width= 0.5\textwidth]{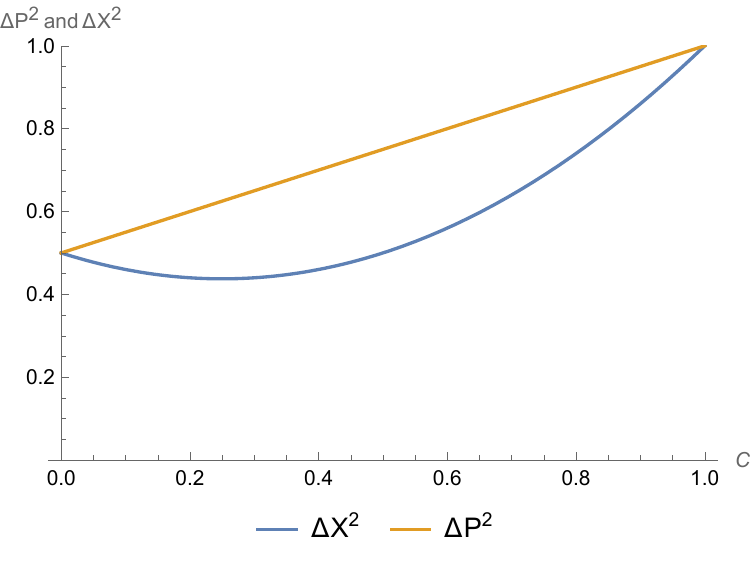}
  \caption{Plot of Quadrature squeezing for dispersion $\Delta X^2$ and $\Delta P^2$ versus concurrence for angle $\phi =0, \pi$}
  \label{fig:concurrence}
\end{figure}
\end{center}
\newpage

\subsection{Golden Uncertainty Relation and Fibonacci Numbers}

In above calculations (\ref{criticaldispersion}) we have seen that dispersion $\left( \Delta {P} \right)^{2} = \frac{5}{8} = \frac{F_5}{F_6}$ is equal to the ratio of two Fibonacci numbers. Moreover,  for $C=\frac{1}{4}$ and $\phi =0$, from (\ref{X2}), (\ref{P2}) we have $(\Delta X)^2 = \frac{7}{16}$,  $(\Delta P)^2 = \frac{5}{8}$, while for $\phi = \frac{\pi}{2}$, it is $(\Delta X)^2 = \frac{5}{8}$,  $(\Delta P)^2 = \frac{7}{16}$.
In fact, Fibonacci numbers and Golden Ratio can be involved to uncertainty relations for super-coherent states as follows.
For maximally random states (\ref{maxrandom}) with $C = \frac{1}{2}$, located on the equator of the super Bloch sphere,
the uncertainty relation is
\begin{equation}
\Delta X \Delta P = \frac{1}{8} \sqrt{24 + \sin^2 2\phi}. \label{XPMAX}
\end{equation}
 For angle $\phi = \frac{\pi}{4}$ it gives ratio of two Fibonacci numbers.
\begin{equation}
\Delta X \Delta P = \frac{5}{8} = \frac{F_5}{F_6}. \label{F5F6}
\end{equation}
In addition we notice that  the minimal uncertainty $\frac{1}{2} = \frac{F_2}{F_3}$ and the maximum uncertainty
$1 = \frac{F_1}{F_2}$ also represent the ratio of two Fibonacci numbers. In all these cases the uncertainty is equal to the ratio $\frac{F_n}{F_{n+1}}$.
The inverse ratio of it
\begin{equation}
\varphi_n = \frac{F_{n+1}}{F_n},
\end{equation}
represents the Golden sequence, satisfying equation
\begin{equation}
\varphi_n = 1 + \frac{1}{\varphi_{n-1}},
\end{equation}
and having  the Golden ratio as the limit $\varphi_n \rightarrow \varphi$, when $n \rightarrow \infty$. This suggests that for supersymmetric coherent states exist the sequence of
uncertainties, equal $1/{\varphi_{n}}$ for any positive $n$,  giving in the limit $n \rightarrow \infty$ the Golden Ratio uncertainty
\begin{equation}
\Delta X \Delta P = \frac{1}{\varphi} = \frac{2}{1 + \sqrt{5}}. \label{goldenuncertainty}
\end{equation}
To determine the golden sequence for any $n$,  we fix  the angle $\phi = \frac{\pi}{4}$, then dispersions are equal $\Delta X_n = \Delta P_n$   and due to (\ref{square}) they can be choosen as
\begin{equation}
\Delta X_n \Delta P_n =\frac{F_n}{F_{n+1}} = \frac{1+ C^2_n}{2}. \label{Fsequence}
\end{equation}
This implies the infinite sequence of concurrences, determined by equation
\begin{equation}
C_n^2 +1 = 2 \frac{F_n}{F_{n+1}}.
\end{equation}
Using properties of Fibonacci numbers, $F_{n+1} = F_n + F_{n-1}$, it can be simplified as
\begin{equation}
C^2_n = \frac{F_{n-2}}{F_{n+1}},
\end{equation}
so that
\begin{equation}
C_n = \sqrt{\frac{F_{n-2}}{F_{n+1}}}.
\end{equation}
In particular case $n=5$ it gives value $C = \frac{1}{2}$ considered above (\ref{F5F6}), and for $n = 1$ and $n =2$, the maximal and minimum uncertainties, $1$ and $\frac{1}{2}$, correspondingly.
  For successive $n$ and $n+1$ terms,  dispersions in $X$
\begin{equation}
(\Delta X_n)^2 = \frac{F_n}{F_{n+1}}, \hskip1cm (\Delta X_{n+1})^2 = \frac{F_{n+1}}{F_{n+2}},
\end{equation}
give the ratio of Fibonacci numbers
\begin{equation}
\frac{(\Delta X_{n+1})^2}{(\Delta X_n)^2} = \frac{F_{n+2}}{F_n}.
\end{equation}
In the limit $n \rightarrow \infty$ it takes the form of the Golden Ratio
\begin{equation}
lim_{n \rightarrow \infty}\frac{\Delta X_{n+1}}{\Delta X_n} = \varphi = \frac{1 + \sqrt{5}}{2}.
\end{equation}
This formula shows how the Golden Ratio naturally  appears in supersymmetric quantum oscillator, as a ratio of the uncertainties.

The set of super-coherent states corresponding to the golden sequence of  uncertainties (\ref{Fsequence}) is
\begin{equation}
 | \alpha,
\sqrt{\frac{F_{n-2}}{F_{n+1}}}, \frac{\pi}{4} \rangle_{L_\pm} = \sqrt{1 - \sqrt{\frac{F_{n-2}}{F_{n+1}}}} \,|\alpha, \Psi_0 \rangle +
\frac{1+i}{\sqrt{2}} \left(\frac{F_{n-2}}{F_{n+1}} \right)^{1/4}|\alpha, L_\pm \rangle.
\end{equation}
In the limit $n \rightarrow \infty$,
the concurrence $C_n$ is represented by the
Golden Ratio $C_\infty = \varphi^{-3/2}$ and the sequence of corresponding states converges
 to the Golden super-coherent state,
\begin{equation}
 | \alpha, \frac{1}{\varphi^{3/2}}, \frac{\pi}{4} \rangle_{L_\pm} = \sqrt{1 - \frac{1}{\varphi^{3/2}}} \,|\alpha, \Psi_0 \rangle + \frac{1+i}{\sqrt{2} \varphi^{3/4}} |\alpha, L_\pm \rangle.
\end{equation}
For this state we have the Golden Uncertainty relation (\ref{goldenuncertainty}) in the form
\begin{equation}
\Delta X \Delta P = \frac{\hbar}{\varphi}
\end{equation}
(here we recovered the Planck constant). The relation determines the Golden proportion
\begin{equation}
\varphi = \frac{\hbar}{\Delta x \Delta p}
\end{equation}
in the phase plane cells, as ratio of Plank constant with area of the cell.
Moreover, the uncertainty value $h/2\pi \varphi$ corresponds to the length of the circle $2 \pi \varphi$ with radius
$r = \varphi$. Inversion of this circle in the unit one gives the circle with radius $1/\varphi$ and the length $2\pi/\varphi$, which determines the Golden Angle.
This angle appears in the theory of sunflowers \cite{sunflower} as efficiency model of sunflowers packing, and it would be interesting to see how it can be combined with phase space structure in quantum mechanics.

\section{Conclusions}

In present paper by using four fermion-boson Bell states, the super-coherent states were introduced and studied in terms of entanglement
and the super-qubit states. Similarly to non-classical PAC states,
non-classicality of our supersymmetric coherent states can be described by Mandel  $Q = \frac{\langle {\cal N}^2 \rangle - \langle {\cal N} \rangle^2}{\langle {\cal N} \rangle} - 1$
parameter, calculated for super-number operator ${\cal N}$.
For the first pair of coherent super-Bell states $|\alpha, L_\pm \rangle$ with $C=1$
we get
\begin{equation}
Q = \frac{|\alpha|^2 -1}{|\alpha|^2 +1}.
\end{equation}
Depending on values of $\alpha$ it shows different statistics. Non-classical statistics occurs for $Q < 0$, and for unit disc $|\alpha| < 1$ it is Sub-Poissonian type, while on the unit circle $|\alpha| =1$,  the Poissonian one. More details on this statistics for other states will be published in separate paper.

If instead of maximally entangled Bell states we use generic one superparticle state  $|1, \theta_1, \phi_1\rangle$ , then the concurrence takes the form
\begin{equation}
C = \sin^2\frac{\theta}{2} \,\sin \theta_1,
\end{equation}
which provides transition from entanglement for super-coherent state (corresponding to $\theta_1 = \pi/2$), to entanglement of the supe-number state (when $\theta =\pi$).

As potential applications of our results the Jaynes-Cummings model for extremely strong coupling is a natural candidate \cite{HussinNieto}.
One more interesting application is related with
 optical supersymmetry, which  as
 shown could provide a versatile platform in synthesizing a new class of optical
structures with desired properties and functionality \cite{SUSYoptics}.

\section{Acknowledgements} This work was  supported by Izmir Institute of Technology BAP project grant 2022IYTE-1-0002.


\begin{thebibliography}{99}

\bibitem[1]{klauder} Klauder, J. R. and Skagerstain, B. S., 1985.\textit{Coherent States-Applications in Physics and Mathematical Physics.} World Scientific.


\bibitem[2]{Agarwal} Agarwal G.S. and Tara K., 1991. Nonclassical properties of states generated by the excitations on a coherent state . \textit{Phys. Rev. A}, Vol. 43, 492-497.

\bibitem[3]{Francis} Francis J. T. and Tame M. S., 2020. Photon-added coherent states using the continuous-mode formalism . \textit{Phys. Rev. A}, Vol. 102, 043709.

\bibitem[4]{Zavatta} Zavatta A., Viciani S. and Bellini M., 2004. Quantum-to-Classical Transition with Single-Photon-Added Coherent States of Light. \textit{Science}, Vol. 306, 660.

\bibitem[5]{aragone} Aragone, C. and Zypman F.,1986. Supercoherent states. \textit{J. Phys. A: Mathematical and General.}, Vol. 19, No. 12, pp.2267-2279.

\bibitem[6]{Cooper} Cooper, F., Khare, A. and Sukhatme, U., 2001. \textit{Supersymmetry in  Quantum Mechanics}. World Scientific.

\bibitem[7]{Hussin} Berube-Lauziere, Y. and Hussin V., 1993. Comments of the definitions of coherent states for the SUSY
harmonic oscillator. \textit{J. Phys. A: Mathematical and General.}, Vol. 26, No. 12, pp.6271-6275.


\bibitem[8]{kornbluth}Kornbluth, M. and Zypman, F., 2013. Uncertainties of coherent states for a generalized supersymmetric annihilation operator. \textit{Journal of Mathematical Physics.}, Vol. 54, No. 1.

\bibitem[9]{Fatyga} Fatyga, B.W. et al., 1991. Supercoherent states. \textit{Physical Review D}, Vol. 43, No. 4, pp.1403-1412.


\bibitem[10]{Nieto} Nieto, M.M., 1991. Physical interpretation of supercoherent states and their associated Grassman numbers .




\bibitem[11]{orszag} Orszag, M. and Salamo, S., 1988. Squeezing and minimum uncertainty states in the
supersymmetric harmonic oscillator. \textit{J. Phys. A: Math. Gen.}, Vol. 21, LlOS9-LlO64.



\bibitem[12]{Zypman} Zypman, F.R., 2015. Supersymmetric Displaced Number States. \textit{Symmetry}, Vol. 7, 1017-1027.



\bibitem[13]{Iliyeva} Ilieva N., Narnhofer H. and Thirring W., 2004. Finite supersymmetry transformations. \textit{Eur.Phys.J.C}, Vol. 35, 119-127.

\bibitem[14]{Laba} Laba, H.P., and Tkachuk, V.M., 2020. Entangled states in supersymmetric quantum mechanics . \textit{Modern Phys. Lett. A}, Vol. 35, No. 34, 2050282.


\bibitem[15]{Motamed} Motamedinasab, A., Afshar, D. and Jafarpour, M., 2018. Entanglement and non-classical properties of generalized
supercoherent states. \textit{Optik}, Vol. 157, 1166-1176.

\bibitem[16]{Jonsson} Jonsson, R.H., Hackl, L. and Roychowdhury, R., 2021. Entanglement dualities in supersymmetry. \textit{Phys. Rev. Research }, Vol. 3, 023213.



\bibitem[17]{SUSYoptics} Mohammad-Ali Miri, Matthias Heinrich, Ramy El-Ganainy, and Demetrios N. Christodoulides, 2013. Supersymmetric Optical Structures. \textit{Phys. Rev. Lett.}, Vol. 110, 233902.


\bibitem[18]{Pashaevkocak2} Pashaev, O.K. and Ko\c{c}ak, A., 2019.
Special functions with mod n symmetry and kaleidoscope of quantum coherent states, \textit{Journal of Physics: Conference Series}, Vol. 1194, 012059.

\bibitem[19]{PashaevGurkan} Pashaev, O.K. and Gurkan, N.,  2012. Energy localization in maximally entangled two- and three-qubit phase space.
\textit{New Journal of Physics}, Vol.14, 063007.


\bibitem[20]{Fernandez} Garcia-Munoz, J.D., Fernandez, D.J. and Vergara-Mendez, F., 2023.  Supersymmetric quantum mechanics, multiphoton algebras and coherent states.
\textit{Physica Scripta}, Vol. 98, No. 10, 105243.


\bibitem[21]{sunflower} Newell, A.C. and Pennybacker, M. ,2013. Fibonacci patterns: common or rare?. \textit{Procedia IUTAM}, Vol. 9, 86-109.

\bibitem[22]{HussinNieto} Hussin, V. and Nieto, L.M., 2005. Ladder operators and coherent states for the Jaynes-Cummings model in the rotating-wave approximation. \textit{Journal of Math. Phys.}, Vol. 46,
122102.

\bibitem[23]{ParlakPashaev} Parlakgorur, T. and  Pashaev, O.K., 2019. Apollonius representation and complex geometry of entangled qubit states, \textit{Journal of Physics: Conference Series}, Vol. 1194,
012086.



\bibitem[24] {louisell} Louisell, W.H., 1964. \textit{Radiation and Noise in Quantum Electronics.} McGraw-Hill Book Company.

\bibitem[25]{LE} Fabrizio Buscemi, Paolo Bordone and Andrea Bertoni, 2007. Linear entropy as an entanglement measure in two-fermion systems. \textit{Phys. Rev. A}, Vol. 75, 032301.


\bibitem[26] {benenti} Benenti, G. , Casati, G. and Strini, G., 2004.  \textit{Principles of Quantum Computation and Information, Vol.1: Basic Concepts.} World Scientific.

\bibitem[27] {dodonov} Dodonov, V. V., Malkin, I. A. and Man'ko, V. I., 1974. Even and odd coherent states and excitations of a singular oscillator. \textit{Physica}, Vol. 72, No.3, pp. 597-615.

\bibitem[28]{glauber} Glauber, R. J., 1963. Coherent states and incoherent states of the radiation field. \textit{Physical Review.},Vol 131, No. 6, pp 2766-2788.

\bibitem[29]{afshar} Afshar, D. , Motamedinasab, A. ,Anbaraki, A. and Jafarpour, M.,2016. Even and odd coherent states of supersymmetric harmonic oscillators and their nonclassical properties. \textit{International Journal of Modern Physics B.}, Vol. 30, No. 7.



\end{thebibliography}
\end{document}